\documentclass[aps,prb,twocolumn,showpacs,
twoside,superscriptaddress,nofootinbib,amsmath,amssymb ]{revtex4-2}

\usepackage{graphicx}
\usepackage{dcolumn}
\usepackage{bm}
\usepackage{tikz}
\usepackage{circuitikz}
\usepackage{hyperref}
\usepackage[mathlines]{lineno}
\usepackage{float}
\usepackage{subfloat}

\begin{document}

\preprint{APS/123-QED}

\title{Long-range Ising spins models emerging from frustrated Josephson junctions arrays with topological constraints
}

\author{Oliver Neyenhuys}
\author{Mikhail V. Fistul}
\author{Ilya M. Eremin}
\affiliation{Theoretische Physik III, Ruhr-Universit\"at Bochum, Bochum 44801, Germany}

\date{\today}

\begin{abstract}
Geometrical frustration in correlated systems can give rise to a plethora of novel ordered states and intriguing phases. Here, we analyze theoretically vertex-sharing frustrated Kagome lattice of Josephson junctions and identify various classical and quantum phases. The frustration is provided by periodically arranged $0$- and $\pi$- Josephson junctions.  In the frustrated regime the macroscopic phases are composed of different patterns of vortex/antivortex penetrating each basic element of the Kagome lattice, i.e., a superconducting triangle interrupted by three Josephson junctions. We obtain that numerous topological constraints, related to the flux quantization in any  hexagon loop,  
lead to highly anisotropic and long-range interaction between well separated vortices (antivortices). Taking into account this interaction and a possibility of macroscopic "tunneling" between vortex and antivortex in  single superconducting triangles we derive an effective Ising-type spin Hamiltonian with strongly anisotropic long-range interaction. In the classically frustrated regime we calculate numerically the temperature-dependent spatially averaged spins polarization, $\overline{m}(T)$, characterizing the crossover between the ordered and disordered vortex/antivortex states. In the coherent quantum regime we analyze the lifting of the degeneracy of the ground state and the appearance of the highly entangled states. 
\end{abstract}

\maketitle

\section{Introduction}
The collective behavior of the low-energy magnetic excitations crucially depend on the geometry of the lattice they inhabit. For example, antiferromagnetically interacting spins on a square lattice form a N\'eel order with antialigned neighbours. At the same time, their mutual antiparallel alignment cannot be satisfied on a triangular or kagome lattices, which are the most typical models, which feature geometric frustration and yield non-trivial spin order \cite{anderson1978concept,moessner2006geometrical,schroder2005competing,balents2010spin,baniodeh2018high,han2012fractionalized,song2023tensor}. 
The frustration can also be provided by the competition of interactions of alternating signs of the interactions \cite{anderson1978concept,nisoli2013colloquium}, e.g., the ferromagnetic and anti-ferromagnetic ones in addition to a special geometry of the lattices. Typical consequence of the frustration is the highly degenerated ground state, a large amount of low-lying metastable states and long relaxation times at low temperatures \cite{moessner2006geometrical,balents2010spin,mahmoudian2015glassy}. 

Apart from the natural solid state systems demonstrating rich plethora of interesting physics behavior due to underlying frustration like  is found in iron-based  superconductors \cite{gong2017possible,fernandes2014drives}, frustrated ferromagnetic chains \cite{zhitomirsky2010magnon}, Kagome magnets \cite{Yan2011,han2012fractionalized,yamada2016first,messio2012kagome,Fujihala2020,Teng2023} and superconductors \cite{neupert2022charge,feng2021chiral,yin2022topological,Yang2023} a special attention had attracted artificially prepared systems such as 
trapped ions simulators~\cite{britton2012engineered},  photonic crystals~\cite{weimann2016transport,vicencio2015observation}, two-dimensional arrays of Rydberg atoms\cite{Semeghini2021,Giudici2022,Yan2023}, anisotropic optical lattices \cite{Xu2023} and Josephson junctions networks~\cite{rzchowski1997phase,pop2008measurement,NoriKitaev2010,johnson2011quantum,king2018observation,king2021scaling} due to a more efficient way to tune the frustration parameter.

The latter system coined as frustrated Josephson junction arrays (\textit{f-JJA}s) is of a special
interest since the current technology allows to form \textit{f-JJA}s of various geometry and size as well as 
 to tune the frustration  by an externally applied magnetic field \cite{pop2008measurement,douccot2002pairing,cataudella2003glassy}. Furthermore, the physics of \textit{f-JJA}s can be mapped into different non-integrable quasi-$1D$/$2D$ Ising or $X$-$Y$ spins models, and therefore, such arrays can provide a feasible experimental platform to establish {\it analog quantum simulations} in the fields of quantum chemistry, quantum biology and low-dimensional material science \cite{daley2022practical,buluta2009quantum}. 

It is known that the \textit{f-JAA}s display the non-frustrated and the frustrated regimes 
characterized by the unique and highly degenerated ground states, accordingly. In the frustration regime a plenty of complex ground states such as the checkerboard and ribbon distribution of vortices~\cite{rzchowski1997phase,caputo2001resonances}, strip phases~\cite{valdez2005superconductivity} and so on, and sharp transitions between these magnetic patterns as the external magnetic field varies, were observed in \textit{f-JJA}s on square and triangular lattices. 

A special type of \textit{f-JJA}s is 
\textit{vertex-sharing} lattices in which an each site is shared between two neighboring triangles, e.g.,  quasi-$1D$ sawtooth and diamond chains \cite{rizzi20064,pop2008measurement,douccot2002pairing,andreanov2019resonant}, and two-dimensional Kagome lattice \cite{song2023tensor,andreanov2020frustration}. In the frustration regime of such \textit{f-JJA}s the  vortex/antivortex penetrates each single superconducting triangle, and various distributions of vortices/antivortices can be realized. The vortex (antivortex) states correspond to anticlockwise (clockwise) persistent currents flowing in a single triangle. 

The classical frustrated regime of saw-tooth and diamond chains of Josephson junctions has been previously theoretically studied in Ref. \cite{andreanov2019resonant} where the \textit{disordered} state of vortices/antivortices was obtained. The lack of ordering in the distribution of vortices/antivortices in such quasi-$1D$ \textit{f-JJA}s was due to the absence of interaction between vortices/antivortices of different cells. At the same time, for \textit{f-JJA}s based on the Kagome lattice the highly anisotropic distributions of vortices/antivortices forming the ground state has been also predicted \cite{andreanov2020frustration}. 
What, however, remains unclear is what type of the interaction can lead to the formation of such ordered anisotropic patterns and how the order-disorder phase transition in vortex patterns occurs? Moreover, at low temperatures one can expect due to charging effects an intriguing interplay between the quantum superposition of classical vortex/antivortex states  and the interaction of well separated vortices (antivortices).

In this manuscript we address these questions performing a systematic theoretical study of classical and coherent quantum collective states occurring in an exemplary two-dimensional vertex-sharing frustrated Kagome lattice of Josephson junctions. In a complete analogy with the magnetic systems where the ferromagnetic and the antiferromagnetic couplings coexist, we introduce the frustration as a periodic alternation of $0$- and $\pi$-Josephson junctions. The $\pi$-Josephson junctions can be fabricated on a basis of various multi-junctions SQUIDS in externally applied magnetic field~\cite{rzchowski1997phase,pop2008measurement,johnson2011quantum,king2018observation,orlando1999superconducting,jung2014multistability,shulga2018magnetically,king2021scaling}, superconductor-ferromagnet-superconductor junctions ~\cite{feofanov2010implementation}, different facets of grain boundaries of high temperature superconductors~\cite{hilgenkamp2008pi} or Josephson junctions between two-bands superconductors~\cite{dias2011frustrated}.  
By making use of this model we show explicitly that in the frustration regime the vortices/antivortices penetrate a basic element of the Kagome lattice, i.e., a superconducting triangle interrupted by three (two-$0$ and one-$\pi$) Josephson junctions. 
The observed collective states of vortices/antivortices  are determined by numerous \textit{topological constraints} related to the flux quantization in any hexagon loops leading to an effective interaction between vortices (antivortices) of different triangles.



The paper is organized as follows: In Section II we present the electrodynamic model and the general approach allowing one to quantitatively analyze collective classical and quantum states arising in the Kagome lattice of frustrated Josephson junctions with numerous topological constraints occurring in such arrays.
In Section III we study the low-lying states of a single building block of vertex-sharing frustrated Josephson junction arrays \textit{f-JJA}s, i.e., three $\pi/0$ Josephson junctions incorporated in a single superconducting loop. In the frustrated regime we arrive at a single spin model in which two basis spin states correspond to (counter)clockwise persistent currents or the penetration of vortex/antivortex. 
In Section IV our analysis is extended to the Kagome lattice of frustrated Josephson junctions for which we derive an effective $(2D+1)$ Ising spin model with a long-ranged and spatially anisotropic interaction between well separated spins. 
In Section V and VI we analyze the classical and the coherent quantum frustrated regimes and the corresponding phases. Section VII provides conclusions. The details of the calculation of the partition function of interacting spins $Z$ and an explicit spatial dependence of the interaction strength in the infinite Kagome lattice will be presented in Appendices $A$ and $B$, respectively.

\section{\label{sec:model} Model, general approach and topological constraints}

\begin{figure}
    \centering
    \includegraphics[width=0.95\columnwidth]{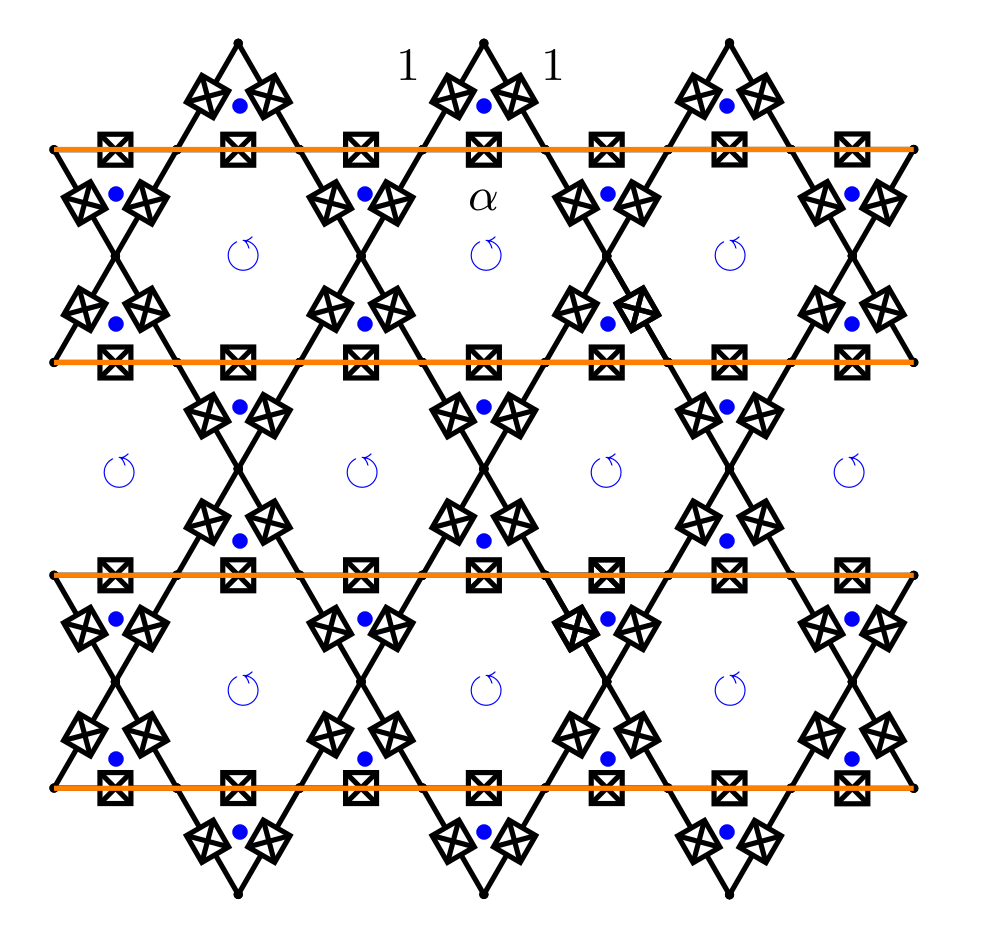}
    \caption{(color online)Schematics of the frustrated Kagome lattice of Josephson junctions. The $\pi$- and the $0$- Josephson junctions are indicated by orange and black lines, respectively. Two kinds of closed loops, i.e., the triangles (blue dots) and hexagons (blue circles), are shown. }
  \label{fig:schematic}  
\end{figure}

We consider a  vertex-sharing $2D$ Kagome lattice of superconducting nodes (islands) in which the adjacent nodes are connected by Josephson junctions as schematically shown in Fig.\ref{fig:schematic}. The frustration is induced by the special periodic  arrangement of $0$- and $\pi$ -Josephson junctions. In particular, the Josephson junction connecting the superconducting $i$ and $j$ nodes is characterized by two physical parameters, the Josephson coupling energy, $\alpha_{ij}E_J$, and the charging energy, $E_C/|\alpha_{ij}|$, where $E_J=\hbar I_c/(2e)$ and $E_C=e^2/(2C)$ are determined by the critical current, $I_c$, and the capacitance $C$, respectively. The parameters $\alpha_{ij}$ were chosen as 
$\alpha_{ij}=\alpha$ for all horizontal links (orange lines in Fig. \ref{fig:schematic})
 and $\alpha_{ij}=1$ for all other links. The parameter $\alpha$ is then varied as $-1\leq \alpha \leq 1$ and, therefore, $\alpha >0$ ($\alpha <0$) defines $0$ ($\pi$)- Josephson junctions. The parameter $\alpha$ also relates to the commonly introduced frustration parameter $f$ as $f=(1-\alpha)/2$, which varies between $0$ and $1$. 

The classical dynamics of Josephson junctions arrays is determined by the time-dependent Josephson phases, $\varphi_{ij}(t)$. The partition function $Z$ can be expressed via the path integral in the imaginary time-representation:
\begin{equation}\label{partitionfunction1}
   Z=\int e^{-\frac{S_E}{\hbar}}\mathcal{D}[\varphi_{ij}(\tau)],
\end{equation}
where the Euclidean action is given by
\begin{equation}
  S_E=\int_0^{\hbar/(k_BT)} \mathcal{L}\{\varphi_{ij}, \dot \varphi_{ij}, ~i\tau \}  d\tau
\end{equation}
and the Euclidean Lagrangian $\mathcal{L}$ of a Josephson junctions array is written as:
\begin{equation}
{\mathcal{L}}=\sum_{\langle ij \rangle}\frac{|\alpha_{ij}|\hbar^2}{8E_C}\left[\dot{\varphi}_{ij}\right]^2+\alpha_{ij} E_J \left(1-\cos \varphi_{ij} \right).
\label{Lagrangian}
\end{equation}
Here, $\langle ij\rangle$ refers to the two nearest neighbor nodes coupled by a Josephson junction. 
Note, the measure $\mathcal{D}[\varphi_{ij}(\tau)]$
takes into account that the Josephson phases $\varphi_{ij}(\tau)$ are not independent and have to satisfy numerous topological constraints. They originate from the fact that the flux quantization, {\it i.e.} the sum of the $\varphi_{ij}$ along any closed loop of the lattice has to be $2\pi n$, where $n$ is an integer. These constraints are taken into account explicitly in Eq.(\ref{partitionfunction1})  as
\begin{equation} \label{PF-2}
Z=\int e^{-\frac{S_{E}}{\hbar}}\left(\prod_l\delta[C_l(\varphi_{kp})-2\pi n] \right) \prod_{\langle ij \rangle} d\varphi_{ij},
\end{equation}
with
$$
C_{\ell}(\varphi_{pk})=\sum_{ \langle pk\rangle \in \text{closed loop}, \ell }\varphi_{pk} 
$$
and $\ell$ is the constraint's number. 

\section{\label{sec: section III} A building block of \textit{f-JJA}\MakeLowercase{s}: 
frustrated regime}

\begin{figure}
    \includegraphics[width=0.95\columnwidth]{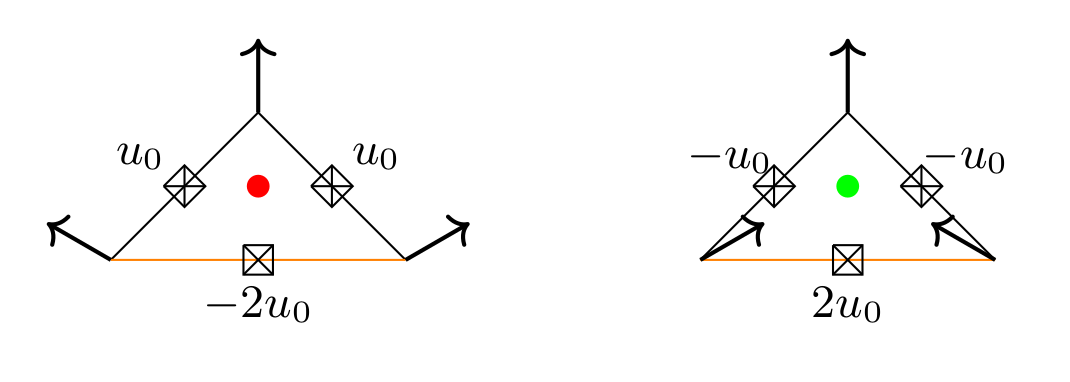}
    \caption{(color online) Schematics of a single superconducting triangle loop interrupted by three Josephson junctions. In the frustrated regime [$\alpha<\alpha_c $ ($f > f_c$)] the classical ground state is doubly degenerate. The penetrating (anti)vortices  are shown by (green)red circles. Corresponding Josephson phases are indicated.}
    \label{fig:single triangle}
\end{figure}

As a starting point of our analysis we now derive the effective Euclidean Lagrangian of the \textit{building block} of the Kagome lattice: a single superconducting triangle interrupted by three Josephson junctions as shown  
in Fig. \ref{fig:single triangle}. 
Such system is characterized by three Josephson phases, $\varphi_{1-3}(\tau)$, which have to satisfy a single constraint, $\varphi_1+\varphi_2+\varphi_3=0$. The corresponding Euclidean Lagrangian of a superconducting triangle then depends on two degrees of freedom, $\varphi_{1,2}$, as 
\begin{eqnarray} \label{Lagr-1}
    \mathcal{L}^\Delta=\frac{\hbar^2}{8E_C}
    \begin{bmatrix}
        \dot{\varphi}_1 & \dot{\varphi}_2
    \end{bmatrix} 
    \begin{bmatrix}
        1+|\alpha| & |\alpha|  \\
        |\alpha|  & 1+|\alpha|
    \end{bmatrix}
    \begin{bmatrix}
        \dot{\varphi}_1 \\
        \dot{\varphi}_2
    \end{bmatrix} +\\
    +E_J[2+\alpha-\cos(\varphi_1)-\cos(\varphi_2)-\alpha \cos(\varphi_1+\varphi_2)].\nonumber
\end{eqnarray} 
Introducing symmetric and anti-symmetric variables  $\varphi_s = (\varphi_1+\varphi_2)/2$ and $\varphi_a = (\varphi_2-\varphi_1)/2$ one can obtain the extrema of the potential energy. If $\alpha< \alpha_c=-0.5$ ($f>f_c=3/4$) the potential energy has \textit{two } equivalent minima as $\varphi_s=\pm u_0$ with 
$u_0=\arccos\left(\dfrac{1}{2|\alpha|} \right)$, and $\varphi_a=0$. These minima are separated by the potential barrier of the height, $E_J(\alpha)=E_J[2(1+\alpha)+1/(2\alpha)]$, which becomes zero at the critical value of $\alpha=\alpha_c$.
Thus, the classical ground state is doubly degenerate (frustrated)  and corresponds to the counterclockwise (a \textit{vortex}) or clockwise (an \textit{antivortex}) persistent currents (see, Fig. \ref{fig:single triangle}). Notice here that the frustrated regime is completely equivalent to the states of a flux qubit at the symmetry point \cite{orlando1999superconducting,jung2014multistability}.

Restricting ourselves to a study of low-lying excitations 
we now use the following approximations.
First, we neglect high frequency oscillations of the asymmetric Josephson phase, $\varphi_a$. Second, the exact dependence of the potential energy on $\varphi_s$ (Eq. (\ref{Lagr-1})) is approximated by harmonic potentials around the two classical minima. The effective "classical" spin degree of freedom $\sigma=\pm 1$ is introduced to distinguish between the vortex/antivortex states. Then the Lagrangian of a single superconducting triangle is written as 
\begin{equation} \label{Lagrangian-ST-2}
    \mathcal{L}^\Delta= \frac{\hbar^2 \gamma}{4E_C}\dot{\varphi}_s^2+\frac{E_J(\alpha)}{u_0^2}[\varphi_s(\tau)-u_0\sigma(\tau)]^2,
\end{equation}
where $\gamma=1+2|\alpha|$. Such reduced Lagrangian depends on two degrees of freedom: the Josephson phase $\varphi_s(\tau)$ and the spin values, $\sigma (\tau)$. 
Since the Lagrangian (\ref{Lagrangian-ST-2}) depends quadratically on $\varphi_s(\tau)$ one can integrate out this degree of freedom, and obtain the \textit{effective spin model}. Indeed, for a single triangle the partition function $Z^\Delta \{\sigma (\tau)\}$ is calculated by expanding  $\varphi_s(\tau)$ and $\sigma (\tau)$ in the sum over the Matsubara frequencies,  $\omega_m=2\pi k (k_B T)/\hbar$, where $k=0, \pm 1...$, i.e., 
\begin{equation}
    \varphi_s(\tau)=\sum_{\omega_k} \tilde{\varphi}(\omega_k) e^{i\omega_k\tau},
\end{equation}
and 
\begin{equation}
    \sigma (\tau)=\sum_{\omega_k} \tilde{\sigma}(\omega_k) e^{i\omega_k\tau}.
\end{equation}
Integrating over all $\tilde{\varphi}(\omega_k)$ we obtain
\begin{equation}
    Z^\Delta \{\tilde{\sigma}(\omega_k)\} \propto \exp\left(\frac{E_J(\alpha)}{k_B T}\sum_{\omega_k} \frac{\Omega^2}{\omega_k^2+\Omega^2}|\tilde{\sigma} (\omega_k)|^2 
    \right),
\end{equation}
where the characteristic frequency $\Omega$ of small oscillations of $\varphi_s$ is $\Omega=[2/(\hbar u_0)] \sqrt{E_CE_J(\alpha)/\gamma}$. 

\section{\label{sec: 4} Effective Ising spins Hamiltonian}
\begin{figure}
    \includegraphics[width=0.95\columnwidth]{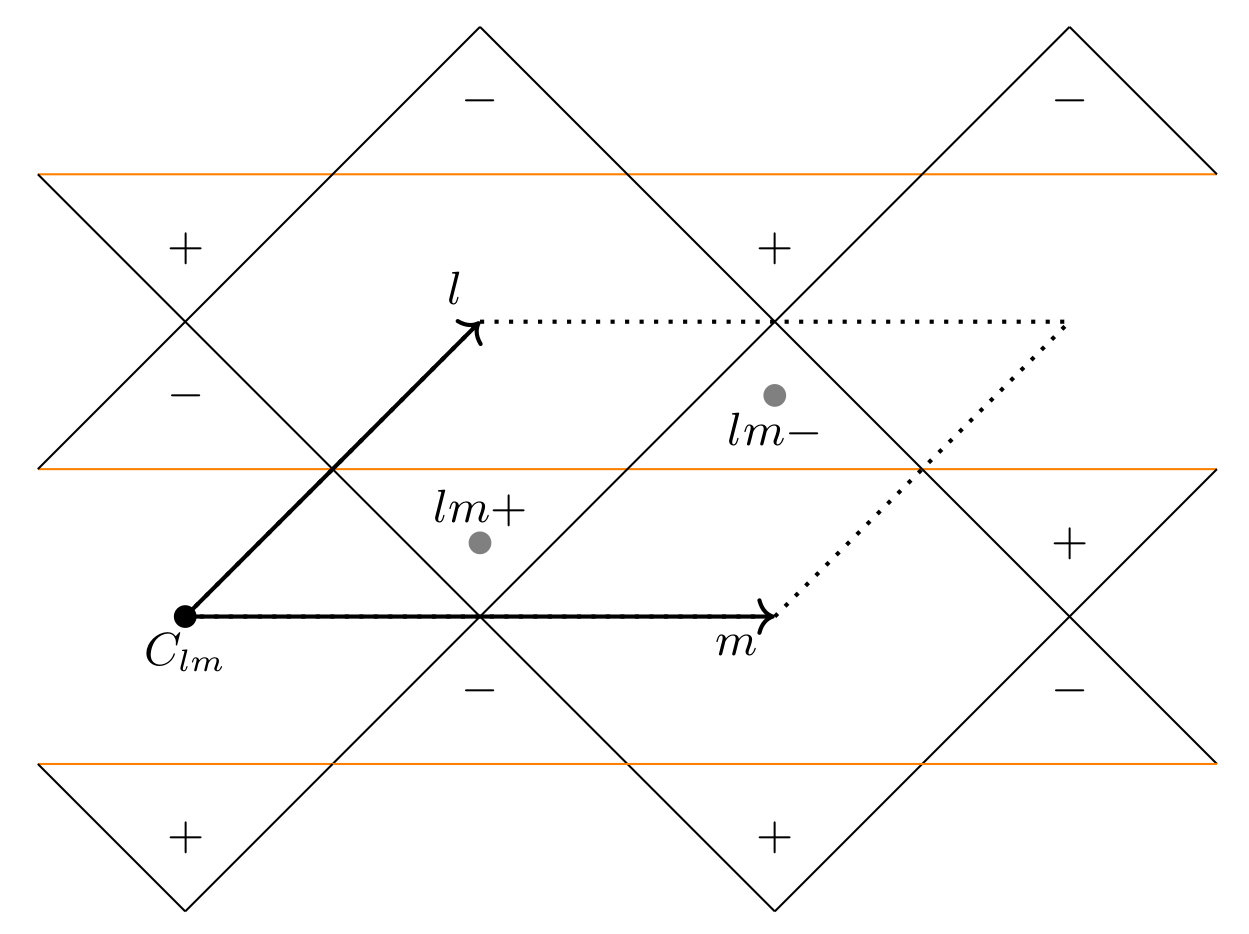}
    \caption{(color online) Representation of the Kagome lattice as the $2D$ periodic lattice of tuples. Each tuple $(l,m)$ labels a rhombus  with sides connecting the nearest hexagon loops centers, and contains a downward ($+$) and upward ($-$) pointing two triangles}
    \label{fig:Kagome latticeEM}
\end{figure}

Next, we extend our analysis 
from a single superconducting triangle to the Kagome lattice. It is convenient to present the Kagome lattice as a periodic repetition in two directions $\ell$ and $m$ of the tuple, $(\ell m)$. Each tuple is a rhombus with the sides connecting hexagon loops centers, and it contains two, a downward ($+$) and an upward ($-$) pointing triangles. Such representation of the Kagome lattice is presented in Fig. \ref{fig:Kagome latticeEM}.

A single superconducting triangle belonging to the tuple, $(\ell m)$, is characterized by two imaginary time dependent degrees of freedom, $\varphi_{\ell m \pm} (\tau)$ and $\sigma_{\ell m \pm} (\tau)$, and the Euclidean Lagrangian is 
\begin{equation} \label{Lagrangian-RKL}
\mathcal{L}=\sum_{\ell m \pm}\mathcal{L}^{\Delta} \left[\varphi_{\ell m \pm}, \dot \varphi_{\ell m \pm}, \sigma_{\ell m \pm} \right ],
\end{equation}
where the Lagrangian of a single superconducting triangle $\mathcal{L}^{\Delta} $ is determined by Eq.  (\ref{Lagrangian-ST-2}). 

\begin{figure}
    \includegraphics[width=0.95\columnwidth]{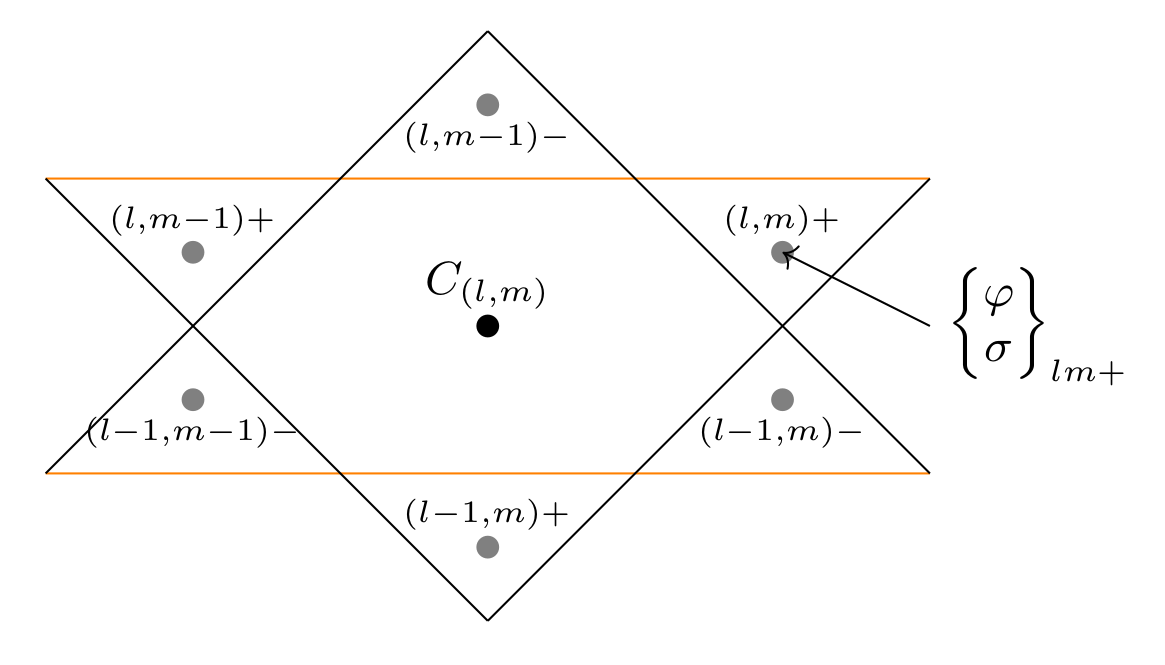}
    \caption{Topological constraint $C_{lm}=0$ for a single hexagon loop. The Josephson phase $\varphi_{\ell, m}$ and spin degree of freedom $\sigma$ are shown.}.
    \label{fig:constraintsfield}
\end{figure}

This Lagrangian has to be accompanied by the topological constraints, $C_{\ell m}=2\pi n$, related to the flux quantization in the hexagon loop $(\ell m )$. Notice that if 
$\alpha$ is not equal to specific values, $-1;-(1/\sqrt{2})$, one can use the constraints with $n=0$ only. The expression of $C_{\ell m}$ depends on the Josephson phases $\varphi_{ij \pm}$ of triangles  surrounding the hexagon loop, $(\ell m)$ (see, Fig. \ref{fig:constraintsfield}). It is written as 
\begin{equation} \label{constraints}
    C_{\ell m}(\tau) = \sum_{ij} \varphi_{ij \pm} (\tau) G_{ij\pm; \ell m},
\end{equation}
where 
\begin{equation}\label{G-function}
\begin{split}
 G_{ij+; \ell m}=-2 \delta_{ij;[(\ell-1) m]}+ \delta_{ij;[\ell (m-1)]}+\delta_{ij;[\ell m]},~~~~~~~~~~~\\
 G_{ij-; \ell m}=-2 \delta_{ij;[\ell (m-1)]}+ \delta_{ij;[(\ell-1) m]}+\delta_{ij;[(\ell-1) (m-1)]}.~~
   \end{split}
\end{equation}
The $\delta_{ij;[\ell m ]}$ is the  Kronecker symbol. 

By making use of the identity
\begin{equation}
\delta\left[C_{\ell m }(\tau)\right]=\int \mathcal{D}[p_{\ell m}(\tau)] \exp \left [-\frac{i}{\hbar}\int_0^{\frac{\hbar }{k_B T}} p_{\ell m} C_{\ell m} d\tau \right ]
\end{equation}
the Eq.(\ref{PF-2}) for the partition function $Z$ is written as
\begin{equation}\label{partitionfunctionKL}
\begin{split}
   Z \propto \int \mathcal{D}[\varphi_{ij \pm}(\tau)]\mathcal{D}[p_{\ell m}(\tau)] \cdot ~~~~~~~~~~~~~\\
   \cdot \exp \left \{-\frac{1}{\hbar}\int_0^{\frac{\hbar}{k_B T}}[\mathcal{L}+ i\sum_{\ell m} p_{\ell m} C_{\ell m} ]d\tau \right \}.
   \end{split}
\end{equation}
Substituting Eqs.(\ref{Lagrangian-RKL})-(\ref{constraints}) in Eq.(\ref{partitionfunctionKL}) and performing all integrals over $\tilde{\varphi}_{\ell m \pm}(\omega_k)$ and $\tilde{p}_{\ell m }(\omega_k)$ (details of straightforward but tedious calculations are presented in the Appendix A)
we obtain the partition function $Z \{\tilde \sigma_{\ell m \pm}(\omega_k)\}$  of the \textit{interacting spins} model as follows:
\begin{equation}\label{effective-spin1}
Z \{\tilde\sigma_{\ell m \pm}(\omega_k)\} \propto \exp\left( -\frac{E_J(\alpha)}{k_B T}\sum_{\omega_k} \frac{\Omega^2}{\omega_k^2+\Omega^2} [\mathcal{F}_{0}+\mathcal{F}_{int}]\right ),
\end{equation}
where the \textit{spatially local } term $\mathcal{F}_{0}$ is expressed as 
\begin{equation}\label{effective-spin2}
\mathcal{F}_0 = -\sum_{\ell m; \pm} |\tilde{\sigma}_{\ell m \pm}(\omega_k)|^2  
\end{equation}
and the interaction term $\mathcal{F}_{int}$ is written as
\begin{equation}\label{effective-spin3}
\begin{split}
\mathcal{F}_{int}=\sum_{\ell m } \sum_{ij} [\tilde{\sigma}_{ij \pm}(\omega_k) G^{\pm\pm}_{ij,\ell m}\tilde{\sigma}_{\ell m \pm}(-\omega_k) + \\
+\tilde{\sigma}_{ij \mp}(\omega_k) G^{\mp\pm}_{ij,\ell m}\tilde{\sigma}_{ \ell m \pm} (\omega_k) ]~~~~~~~~~~~~~~~.
\end{split}
\end{equation}
Here, we introduce the coupling strengths between the spins $\sigma$ of different triangles
\begin{equation} \label{G-pm}
G^{\pm\pm}_{ij,\ell m}=G_{ij \pm,tu}     (G_{+}^\dagger G_{+}+G_{-}^\dagger G_{-})_{tu,vw}^{-1}   G^\dagger_{vw, \ell m \pm}
\end{equation}
and
\begin{equation} \label{G-mp}
    G^{\mp\pm}_{ij, \ell m}=G_{ij \mp,tu}     (G_{+}^\dagger G_{+}+G_{-}^\dagger G_{-})_{tu,vw}^{-1}   G^\dagger_{vw, \ell m \pm}.
\end{equation}
Observe that in all terms of Eqs. (\ref{effective-spin3})-(\ref{G-mp}) the upper (lower) indices were chosen. 

To conclude this Section we notice that the partition function $Z$ of the effective Ising spins model contains two contributions: quantum fluctuations of spins $\sigma$ of individual triangles and the interaction between spins of different triangles.



\section{\label{sec: 5} Classical frustrated regime}

 Here, we present an analysis of the partition function $Z$ and corresponding collective spin (vortices/antivortices) phases in the classically frustrated regime, $k_B T \gg \hbar \Omega$. In this limit the main contribution to the sum over $\omega_k$ in (\ref{effective-spin1}) comes from $\omega_k=0$, and 
 we obtain the classical $2D$ Ising model of interacting spins with the partition function 
\begin{equation}\label{effective-spinClassical}
Z^{cl}\{\sigma_{\ell m \pm}\} \propto \exp\left( -\frac{E_J(\alpha)}{k_B T}
\mathcal{F}^{cl}_{int}\{\sigma_{\ell m \pm}\}
\right ),
\end{equation}
where the Ising-type interaction is given by 
 \begin{equation}\label{effective-interactionCL}
\begin{split}
\mathcal{F}^{cl}_{int}=\sum_{\ell m } \sum_{ij} 
[\sigma_{ij \mp} G^{\mp\pm}_{ij,\ell m}\sigma_{ \ell m \pm}+\sigma_{ij \pm} G^{\pm\pm}_{ij,\ell m}\sigma_{ \ell m \pm} ]~~~~~~~~~~~~~~~.
\end{split}
\end{equation}

One finds that in the classical frustrated regime the observed spins patterns are determined by a single parameter $E_J(\alpha)/k_B T$, and one can identify two distinguished regimes. The first one is realized in the high-temperature limit, $k_B T \gg E_J (\alpha) $, in which all distributions of spins have the same probabilities and we have \textit{disordered } spin configuration.  In the low-temperature regime, $k_B T \ll E_J (\alpha) $ the spins patterns demonstrate a specific ordering determined by the anisotropic and long-ranged interaction strength $\mathcal{F}^{cl}_{int} $.

\begin{figure}
    \includegraphics[width=0.95\columnwidth]{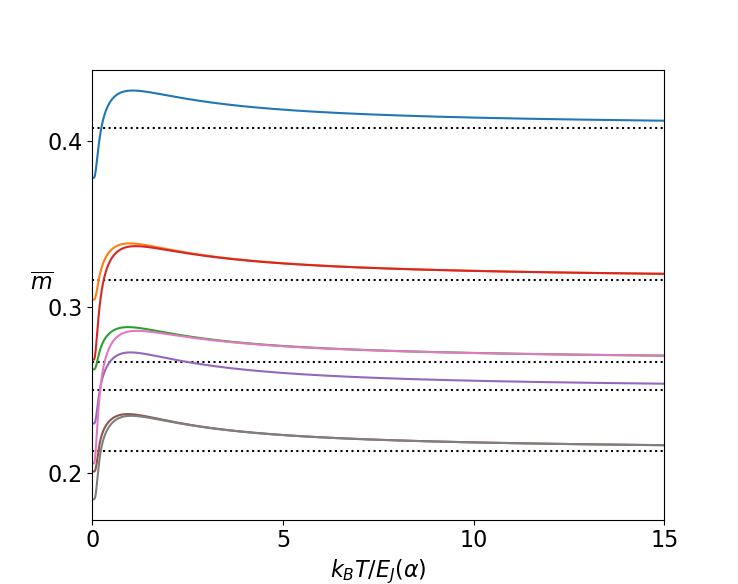}
    \caption{ (color online) Calculated temperature dependence of $\overline{m}(T)$ for the Kagome lattice containing a few plaquettes: $1$, $1 \times 2$ ($2\times 1$), $2\times 2$ and $3\times 2$ ($2\times 3$) 
    are shown. The dotted lines indicate the values of $\overline{m} =1/\sqrt{N}$ in the limit of infinite temperature.} 
    \label{fig:Classicalcrossover}
\end{figure}

To quantitatively characterize the order-disorder phase transition for the Kagome lattice of a \textit{small size }we counted all spin configurations and obtain 
the temperature dependence of a spatially averaged spins polarization $\overline{m}(T)=\sqrt{<M^2>}$, where 
$<M^2>=\sum_M P_M M^2$ and $M= (1/N)(\sum_{ij;\pm} \sigma_{ij;\pm})$. The $P_M$ is the probability of a spin pattern with the spin polarization $M$, and it is determined by the partition function $Z^{cl}$.
Here, $N$ is the total number of triangles. The obtained dependencies of $\overline{m}(T)$ for the  Kagome lattice of different sizes $N$ are presented in the Fig. \ref{fig:Classicalcrossover}.  All curves show the maximum for $k_B T/E_J(\alpha) \simeq 1$ indicating the crossover between the ordered and disordered spins phases.

Since in the high-temperature limit $P_{M=2n-N }\simeq  C^N_n/2^N$, where $n$ is the number of triangles with the spin $\sigma=+1$, we obtain $\overline{m}(T) = 1/\sqrt{N}$ (see dotted lines in Fig. \ref{fig:Classicalcrossover}). In the low-temperature limit as it was shown in Ref. \cite{andreanov2020frustration} the spin patterns become highly anisotropic, and their number  drastically reduces  from $2^N$ to $2^{\sqrt{N}}$ value. Moreover, only few 
spin patterns give non-zero contribution to the spin polarization $m$, and the value of $\overline{m}$ reduces substantially with $N$ in this limit (see, Fig. \ref{fig:Classicalcrossover}).

\begin{figure}[t!]
\includegraphics[width=0.7\columnwidth]{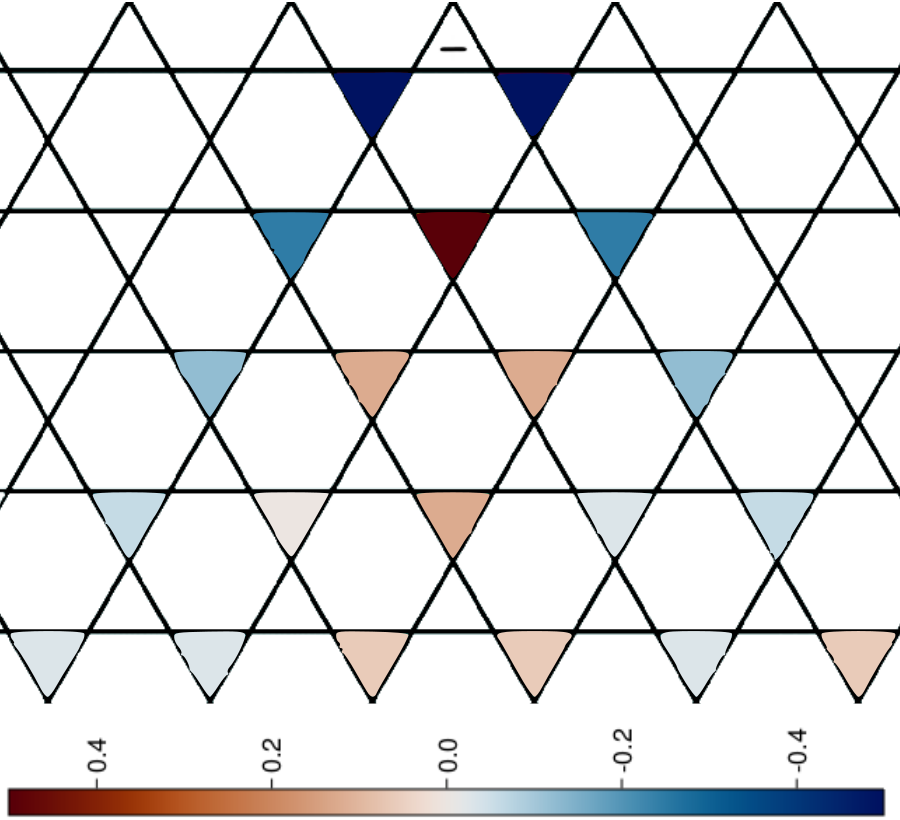}
\Large a)
\bigskip
  \includegraphics[width=0.7\columnwidth]{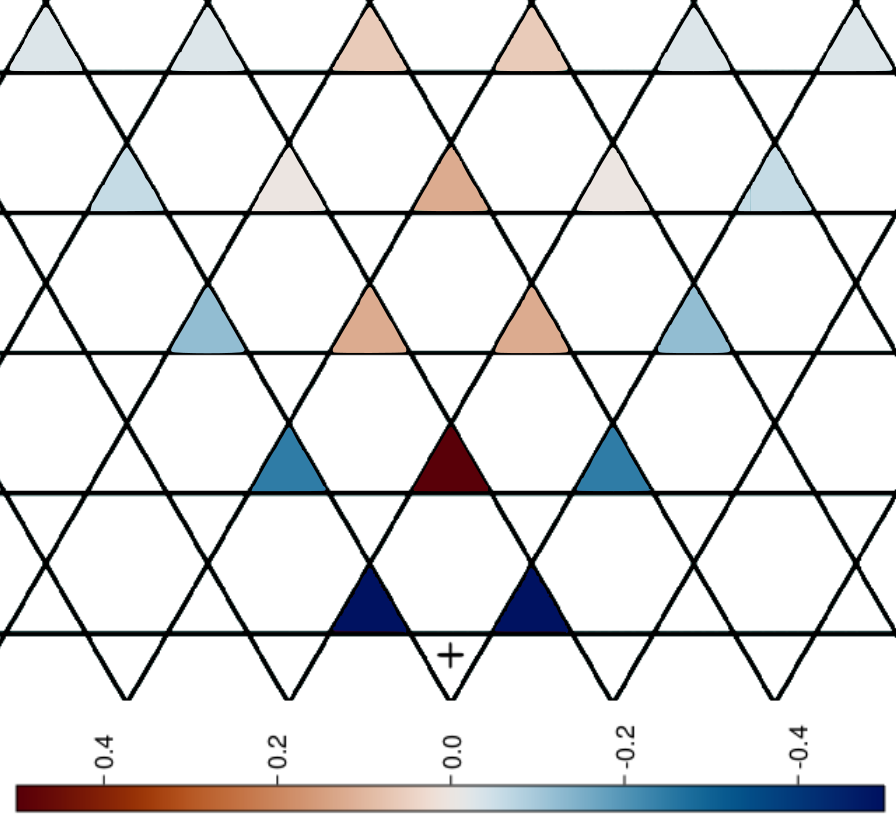}
\Large b)
\caption{ (color online) Calculated two-dimensional color plot of the interaction strength: $G^{-+}_{00; \ell m}$ (a) and $G^{+-}_{\ell m; 00}$ (b) where the downward ($+$) and upward ($-$) pointing triangles $(00)$ are indicated. }
\label{fig:InteractionColorplot-1}
\end{figure}

\begin{figure*}[t!]
\begin{center}
    \includegraphics[width=160pt]{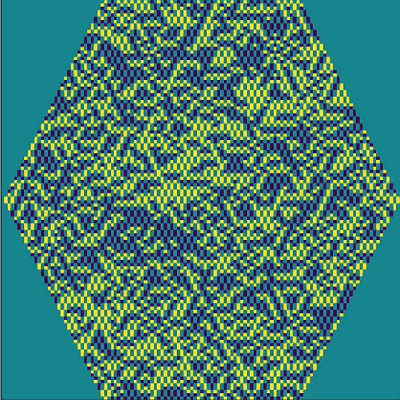}
    \includegraphics[width=160pt]{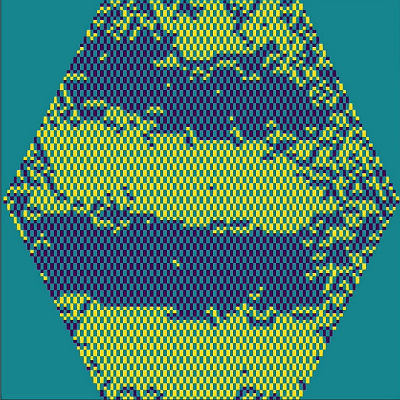}
    \includegraphics[width=160pt]{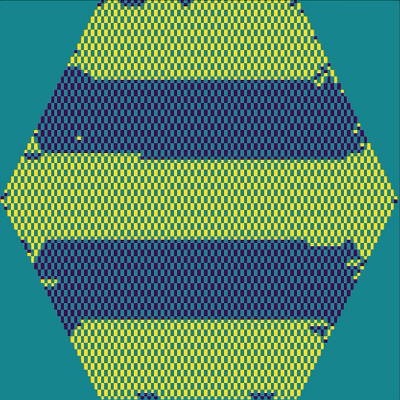}
    \caption{(color online) Calculated typical vortex-antivortex patterns obtained in large $30 \times 30$ frustrated Kagome lattices of Josephson junctions using open boundary conditions for different temperatures: $k_BT/E_J(\alpha)=0.2$ (left); $k_BT/E_J(\alpha)=0.19$ (middle); $k_BT/E_J(\alpha)=0.12$ (right). Quenching from the high-temperature disordered state was used. Dark blue (yellow) rectangles correspond to the vortex (anti-vortex) shared by two vertex-shared triangles (see Fig.4). Light blue rectangles are empty spots.}
   \label{fig:magnetic patterns}
\end{center}
\end{figure*}
   \label{fig:magnetic patterns-2}

As we turn to the \textit{large size} Kagome lattice the interaction strength between spins shows highly anisotropic behavior. The interaction term, $G^{\pm \pm}$, becomes spatially local one, and does not contribute to the classical partition function (\ref{effective-spinClassical}). The interaction strengths $G^{\mp\pm}$ are written as (the details of calculations are presented in Appendix B) 
\begin{equation} \label{eq: InteractionExtended}
\begin{split}
    G_{\ell m; \ell^\prime m^\prime}^{+-}\large|_{m-m^\prime=\ell^\prime-\ell-1}= -\dfrac{1}{2}\dfrac{1}{2^{|\ell^\prime-\ell|}},~~~~~~~~~~~~~~~~~~~~~~~~~~~~~~~~\\
    G_{\ell m; \ell^\prime m^\prime}^{+-}\large|_{m-m^\prime =0}= -\dfrac{1}{2}\dfrac{1}{2^{|\ell-\ell^\prime|}},~~~~~~~~~~~~~~~~~~~~~~~~~~~~~~~~~~~~~~~~~~~~\\
    G_{\ell m; \ell^\prime m^\prime}^{+-}\large|_{m^\prime-m\approx(\ell-\ell^\prime)/2} \propto \dfrac{2\sqrt{2}}{(|\ell-\ell^\prime|)^{1/2}\sqrt{\pi}}~~~~~~~~~~~~~~~~~~~~~~~~~~~~~~~~
    \end{split}
\end{equation}
The two-dimensional color plot of $G_{\vec{\rho} \vec{\rho^\prime}}^{+-(-+)}$ is presented in Fig. \ref{fig:InteractionColorplot-1} shows both algebraic and exponential decay of the interaction strengths in vertical and horizontal directions. Moreover, the interaction 
$G_{00 ; \ell m }^{-+}$ ($G_{\ell m ; 00  }^{+-}$ ) is absent in upper (lower) part of the lattice. 

To observe the thermodynamic phase transition we perform classical Monte-Carlo numerical simulations of the $30 \times 30$ frustrated Kagome lattices of Josephson junctions based on the partition function Eqs. (\ref{effective-spinClassical}) and (\ref{effective-interactionCL}) employing open boundary conditions. In all simulations we started from the disordered (high-temperature) state and cool down up to the temperature $T$. With this procedure we obtain the transition from the disordered vortex-antivortex patterns at $k_B T/E_J(\alpha) \geq 0.2$ to the stripe-type antiferromagnetic order for $k_B T/E_J(\alpha) \leq 0.19$ and below where the ordering is ferromagnetic along $x$-direction and antiferromagnetic along $y$-direction with the period of the stripe of about 11 coupled vertex-shared triangles. Interestingly enough the stripe type AF order sets in despite of the long-range frustrated interaction along $y$-direction and initial 14-fold degeneracy of the ground state for a single Kagome plaquette. At low temperatures we also observed more complex patterns with long domain walls stretched along other symmetry axes (not shown).

\section{\label{sec:6} The coherent quantum regime}

In the limit $k_B T \ll \hbar \Omega$ the quantum fluctuations, and therefore, non-zero Matsubara frequencies in Eq.(\ref{effective-spin1}) start to play an important role. Calculating the sum over the Matsubara frequencies in Eq.(\ref{effective-spin1})
we obtain that a spatially local term $\propto \mathcal{F}_0$ produces the interaction in the imaginary time domain as $\sigma (\tau_i)\sigma (\tau_j)$, where $|\tau_i-\tau_j| \simeq 1/\Omega$. For $E_J(\alpha) \gg \hbar \Omega$ this interaction yields a small amplitude quantum tunneling between the vortex and the antivortex of single triangles. The spatially non-local terms in Eq.(\ref{effective-spin1}) lead to the strong Ising type of interaction described by Eq. (\ref{effective-interactionCL}).
Putting all terms together we obtain the total Hamiltonian in this regime
\begin{equation}\label{Total_Hamiltonian}
\hat H=\hat H_{loc}+ \hat H_{int},
\end{equation}
where 
\begin{equation}\label{Local_Hamiltonian}
\hat H_{loc}=\sum_{\ell m;\pm} \Delta \hat \sigma^x_{\ell m;\pm}
\end{equation}
and
\begin{equation}\label{int_Hamiltonian}
\hat H_{int}=E_J(\alpha) \hat {\mathcal{F}}^{cl}_{int} \{\hat \sigma^z_{\ell m;\pm};\hat \sigma^z_{ij;\pm}\}.
\end{equation}
where $\hat{\mathcal{F}}^{cl}_{int} $ is described by (\ref{effective-interactionCL}), and the tunneling amplitude is $\Delta \simeq \hbar \Omega \exp [-2E_J/(\hbar \Omega)]$.
Thus, the low-temperature quantum regime is described by Eq.(\ref{Total_Hamiltonian}), which corresponds to 
 the seminal model of interacting Ising spins in the transverse magnetic field. 
\begin{figure}
    \centering
    \includegraphics[width=\columnwidth]{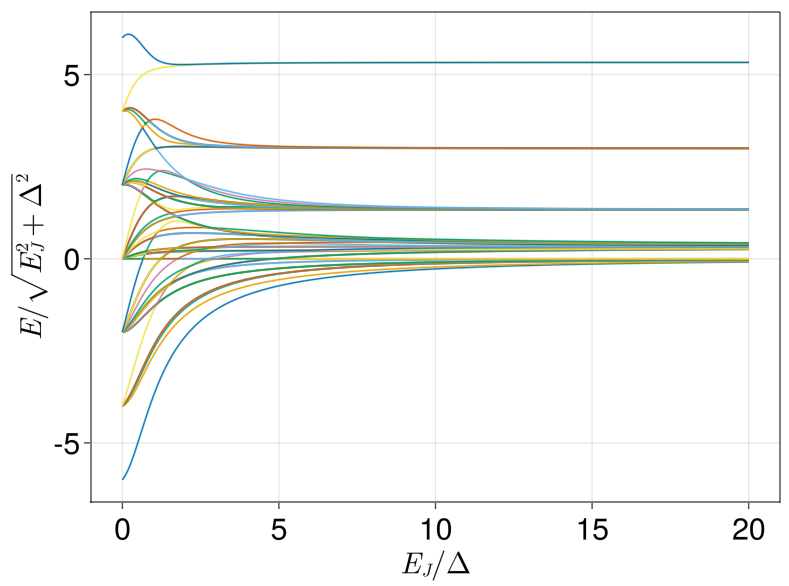}
    \caption{(color online) Calculated energy spectrum as a function of dimensionless parameter $E_J(\alpha)/\Delta$.}
    \label{fig:dispersion}
\end{figure}

The quantum dynamics of such a model is determined by two parameters, $\Delta$ and $E_J(\alpha)$. Carrying out the direct numerical diagonalization we obtain the eigenvalues and eigenfunctions of \textit{all } quantum states in a single plaquette of the Kagome lattice as a function of the ratio $E_J/\Delta$. Such dependence of the normalized eigenspectrum, $E/\sqrt{E_J^2+\Delta^2}$ is presented in Fig. \ref{fig:dispersion}. In the absence of the interaction, $E_J(\alpha) \ll \Delta$ the quantum ground state of each superconducting triangle is the symmetric quantum superposition of vortex and anti-vortex states. The ground state of a whole plaquette is the direct product of these states, i.e., the symmetric quantum superposition of all $64$ classical vortex (antivortex) states, and the quantum entanglement is absent in this regime. Notice also that this regime can be realized as the parameter $\alpha$ is in close vicinity of the critical value $\alpha =-1/2$. 

\begin{figure}
    \centering
    \includegraphics[width=\columnwidth]{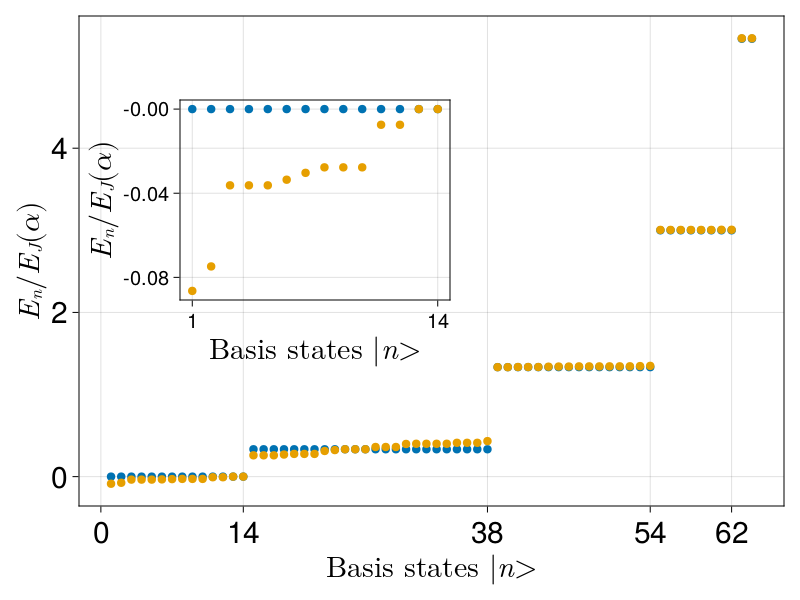}
    \caption{(color online) Calculated energy distribution of different spin patterns for a single plaquette of the Kagome lattice: the classical and the coherent quantum regimes are shown by the blue and the yellow points, respectively. An enlarged area of $14$ low-lying eigenstates is shown in the inset.  The tunneling parameter $\Delta = E_J(\alpha)/20$ was used.}
    \label{fig:energy}
\end{figure}

However, the most interesting case occurs  as the tunneling amplitude $\Delta$ is much smaller than the interaction strength $E_J(\alpha)$. In this limit the eigenspectrum is split into the well-defined bands as shown in Fig. \ref{fig:energy} for the parameter $\Delta=E_J/20$. The lowest band 
contains different quantum superpositions of $14$-fold degenerate classical ground state presented explicitly in \cite{andreanov2020frustration}.

\begin{figure}
    \centering
    \includegraphics[width=\columnwidth]{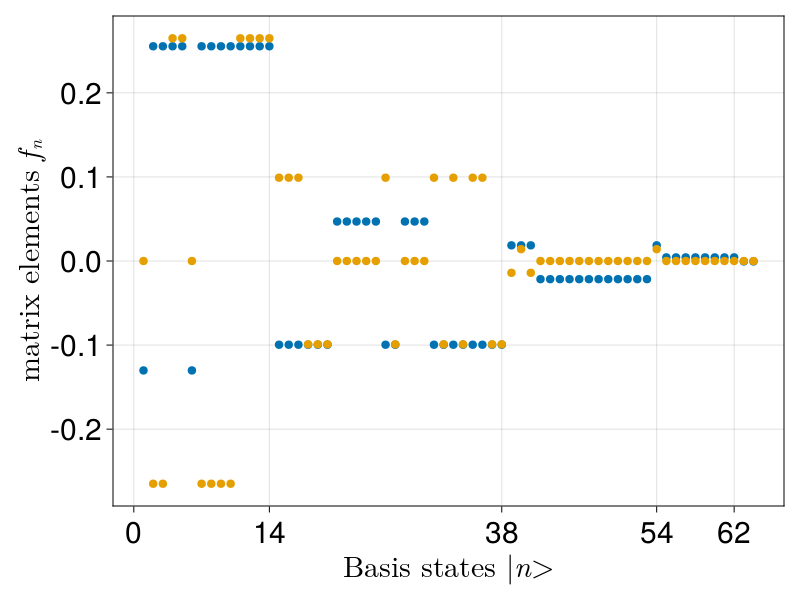}
    \caption{(color online) Calculated wavefunctions  of the ground (blue points) and the first excited (yellow points) states represented by their matrix elements to the basis states $|n>$ of the classical spin interacting Hamiltonian, $\hat H^{cl}$. The tunneling parameter $\Delta = E_J(\alpha)/20 $ was chosen. }
    \label{fig:eigenstates-1}
\end{figure}

\begin{figure}
    \centering
    \includegraphics[width=\columnwidth]{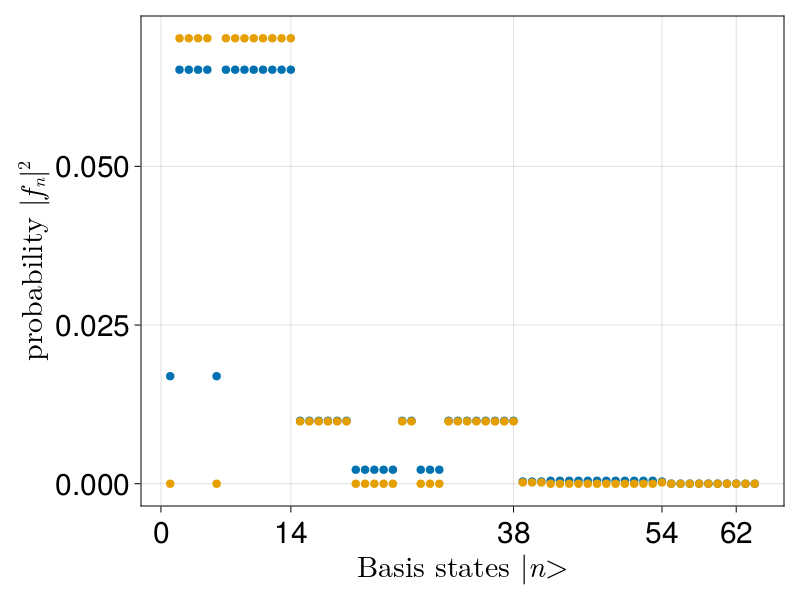}
    \caption{(color online) Calculated ground (blue points) and first excited (yellow points) states represented by their probabilities of the basis states $|n>$ of the classical spin interacting Hamiltonian, $\hat H^{cl}$. The tunneling parameter is $\Delta = E_J(\alpha)/20$.}
    \label{fig:eigenstates-2}
\end{figure}

Furthermore, we obtain that if in the classical regime the ground state is $14$-fold degenerate (blue points in Fig. \ref{fig:energy} ), even a tiny amplitude of the quantum tunneling, $\Delta$, results in the lifting of this classical degeneracy and the corresponding energy distribution among different spin patterns (yellow points in Fig. \ref{fig:energy} ). Moreover, in the quantum regime the wave functions $\Psi$ of the ground and excited states are formed as highly entangled combination of basis (classical) spin patterns, $|n>$. The obtained corresponding matrix elements $f_n=<\Psi|n>$ and probabilities $|f_n|^2$ are shown in Figs. \ref{fig:eigenstates-1}-\ref{fig:eigenstates-2}. One sees that the maximum contribution to the quantum ground and first excited states comes from the $12$ (two highly symmetric vortex (antivortex) patterns are excluded) vortex-antivortex patterns forming the classical ground state.

To conclude this Section we notice that the analysis of spatio-temporal correlations of the quantum ground state of the Hamiltonian (\ref{Total_Hamiltonian})-(\ref{int_Hamiltonian}) in \textit{large} Kagome lattices of Josephson junctions requires using of special methods going beyond direct numerical diagonalization, and will be presented elsewhere.


\section{\label{sec:7} Conclusion}
To conclude, we analyzed theoretically the classical and the quantum phases occurring in a frustrated vertex-sharing $2D$ Kagome lattice of Josephson junctions where
the frustration is provided by a periodic arrangement of $0$- and $\pi$- Josephson junctions.  The frustrated regime is characterized by a highly-degenerated ground state once the frustration parameter $f$ exceeds the critical value $f_c=3/4$. In this regime the (counter)clockwise persistent current flows in a single building block of the Kagome lattice, i.e., a superconducting triangle interrupted by three Josephson junctions (Fig. \ref{fig:single triangle}). These persistent currents are characterized by classical spin values of $\sigma=\pm 1$. 

The quantitative analysis of different patterns of persistent currents (vortices/antivortices)
occurring in the Kagome lattice of Josephson junctions is made by the exact mapping of the initial model of an \textit{f-JJA} onto the effective Ising Hamiltonian of interacting spins. The presence of numerous topological constraints related to the flux quantization in any hexagon loop of the Kagome lattice, results in a long- and short-ranged highly anisotropic interaction between well-separated spins. For the Kagome lattices of a large size the interaction shows a weak algebraic decay  along the vertical axis  and the exponential decay along the horizontal axis (Fig. \ref{fig:InteractionColorplot-1} and Eq. (\ref{eq: InteractionExtended}) ), absence of a non-local interaction between spins of the same pointing triangles. The spatial anisotropy of the vortex/antivortex phases obtained at $T=0$ \cite{andreanov2020frustration} is naturally explained  by the anisotropy of the interaction $G^{+-(-+)}$. 

In the classically frustrated regime, $k_B T \gg \hbar \Omega$, we obtained the order-disorder  phase transition (crossover due to a finite size system) in the patterns of persistent currents. Such crossover is presented in Fig. (\ref{fig:Classicalcrossover}) as the temperature dependent spins polarization, $m(T)$, for the Kagome lattice composed of a few plaquettes. For Kagome lattices of a large size we obtain the disordered vortex-antivortex patterns at high temperatures and stripe-type antiferromagnetic order at low temperatures. 

In the quantum regime we obtain that the macroscopic tunneling between vortices and antivortices in a single superconducting triangle lifts the degeneracy of the classical ground state, and, in the absence of interaction, yields to the symmetric quantum superposition of all $64$ classical states in a single plaquette of the Kagome lattice (see, Fig. \ref{fig:dispersion}). Such quantum ground state demonstrates a zero entanglement. However, as the topological constraints induced interaction is large, i.e., $E_J(\alpha) \gg \Delta$, even a tiny amplitude of the quantum tunneling, $\Delta$, results in the quantum ground state composed of the highly entangled combination of basis (classical) persistent currents patterns (see, Fig. \ref{fig:energy}-\ref{fig:eigenstates-2}).

Finally we notice that frustrated arrays of Josephson junctions arranged in vertex-sharing $2D$ lattices can be used as an ideal physical platform to establish the quantum modelling of generic Ising spin models with constraints \cite{lucas2014ising}, to obtain various collective states in magnetic, optical and superconducting systems. 

\begin{acknowledgments}
 We acknowledge the financial support through the European Union’s Horizon 2020 research and innovation program under grant agreement No 863313 'Supergalax'. We thank Alexei Andreanov for fruitful discussions. 
\end{acknowledgments}


\appendix
\section{Derivation of the partition function of the effective Ising model of interacting spins}
We start from  Eq. 
$(\ref{partitionfunctionKL})$ for the partition function taking into account the topological constraints and the Lagrangians of single triangles 
\begin{equation}
    \begin{split}
   Z \propto \int \mathcal{D}[\varphi_{ij \pm}(\tau)]\mathcal{D}[p_{\ell m}(\tau)] \cdot \\
   \cdot \exp \{-\frac{1}{\hbar}\int_0^{\frac{\hbar}{k_B T}}[\sum_{\ell m \pm}\frac{\hbar^2 \gamma}{4E_C}\dot{\varphi}_{\ell m\pm}^2 \\ +\frac{E_J(\alpha)}{u_0^2}[\varphi_{\ell m\pm}(\tau)-u_0\sigma_{\ell m\pm}(\tau)]^2 \\
   + i\sum_{\ell m} p_{\ell m} \sum_{ij\pm} \varphi_{ij (\tau) \pm} G_{ij\pm; \ell m} ]d\tau \}.
   \end{split}
\end{equation}
Introducing $M=\hbar^2\gamma/2E_C$, exchanging the indices in the second summation and expanding the quadratic term the partition function can be expressed as
\begin{equation}
    \begin{split}
   Z \propto \int \mathcal{D}[\varphi_{ij \pm}(\tau)]\mathcal{D}[p_{\ell m}(\tau)] \cdot \\
   \cdot \exp \{-\frac{M}{2\hbar}\int_0^{\frac{\hbar}{k_B T}}[\sum_{\ell m \pm}\dot{\varphi}^2_{\ell m\pm}+\Omega^2(\varphi_{\ell m\pm}^2+u_0^2)\\
   +\varphi_{\ell m \pm}(-2u_0\Omega^2\sigma_{\ell m\pm} + i\dfrac{2}{M}\sum_{i j} p_{ij} G_{\ell m\pm; ij}) ]d\tau \}.
   \end{split}
\end{equation}
In the next step we switch to the Matsubara representation
\begin{equation}
    \begin{split}
   Z \propto \int \mathcal{D}[\tilde{\varphi}_{ij \pm}(\omega_k)]\mathcal{D}[\tilde{p}_{\ell m}(\omega_k)] \cdot \\
   \cdot \exp \{-\frac{M}{2k_BT}\sum_{\omega_k}[\sum_{\ell m \pm}(\omega_k^2+\Omega^2)\tilde{\varphi}_{\ell m\pm}(\omega_{-k})\tilde{\varphi}_{\ell m\pm}(\omega_{k})\\
   +2\tilde{\varphi}_{\ell m \pm}(\omega_{-k})(-u_0\Omega^2\tilde{\sigma}_{\ell m\pm}(\omega_{k}) + i\dfrac{1}{M}\sum_{i j} \tilde{p}_{ij}(\omega_k) G_{\ell m\pm; ij}) \\
   ]d\tau \},
   \end{split}
\end{equation}
and notice that we have a quadratic form in every $\tilde{\varphi}_{\ell m\pm}(\omega_k)$. Integrating out the quadratic form, we are left with
\begin{equation}
    \begin{split}
   Z \propto \int \mathcal{D}[\tilde{p}_{\ell m}(\omega_k)] \cdot \\
   \cdot \exp \{-\frac{M}{2k_BT}\sum_{\omega_k}[\sum_{\ell m \pm}-\dfrac{u_0^2\Omega^4}{(\omega_k^2+\Omega^2)}\tilde{\sigma}_{\ell m\pm}(\omega_{-k})\tilde{\sigma}_{\ell m\pm}(\omega_{k})\\ 
   + \dfrac{\sum_{ijqr}\tilde{p}_{qr}(\omega_{-k}) G^\dagger_{qr;\ell m\pm} G_{\ell m\pm; ij} \tilde{p}_{ij}(\omega_{k}) }{M^2(\omega_k^2+\Omega^2)}-\Omega^2u_0^2\delta_{k0}\\
   +i\dfrac{u_0\Omega^2}{M(\omega_k^2+\Omega^2)}\sum_{i j}\tilde{\sigma}_{lm\pm}(\omega_{-k}) G_{\ell m\pm; ij} \tilde{p}_{ij}(\omega_{k})\\
   +i\dfrac{u_0\Omega^2}{M(\omega_k^2+\Omega^2)}\sum_{i j} \tilde{p}_{ij}(\omega_{-k})G^\dagger_{ ij;\ell m\pm}\tilde{\sigma}_{lm\pm}(\omega_{k})]\}.
   \end{split}
\end{equation}
Next, we obtain a quadratic form in the constraint fields $\tilde{p}_{\ell m}(\omega_k)$. For that we write the constraint fields and spins as vectors with indexed lattice number $(\ell m)$. We thus introduce the vectors $\vec{\tilde{p}}$,$\vec{\tilde{\sigma}}_+$ and $\vec{\tilde{\sigma}}_-$ as well as the matrices $G_+$ and $G_-$. The partition function can then be written as
\begin{equation}
    \begin{split}
   Z \propto \int \mathcal{D}[\vec{\tilde{p}}(\omega_k)] \\
   \cdot \exp \{-\frac{M}{2k_BT}\sum_{\omega_k}[\sum_{\pm}-\dfrac{u_0^2\Omega^4}{(\omega_k^2+\Omega^2)}\vec{\tilde{\sigma}}_{\pm}(\omega_{-k})\cdot\vec{\tilde{\sigma}}_{\pm}(\omega_{k})\\
   + \dfrac{\vec{\tilde{p}}^T(\omega_{-k}) (G_+^\dagger G_+ + G_-^\dagger G_-) \vec{\tilde{p}}(\omega_{k}) }{M^2(\omega_k^2+\Omega^2)}\\
   +i\dfrac{u_0\Omega^2}{M(\omega_k^2+\Omega^2)}\vec{\tilde{\sigma}}^T_{\pm}(\omega_{-k}) G_\pm \vec{\tilde{p}}(\omega_{k})\\
   +i\dfrac{u_0\Omega^2}{M(\omega_k^2+\Omega^2)}\vec{\tilde{p}}^T(\omega_{-k})G^\dagger_\pm\vec{\tilde{\sigma}}_{\pm}(\omega_{k})]\}.
   \end{split}
\end{equation}
Integrating out the quadratic form over $\vec{p}$ we obtain the final expression
\begin{equation}
    \begin{split}
   Z \propto\\
    \exp \{-\frac{Mu_0^2\Omega^2}{2k_BT}\sum_{\omega_k}\dfrac{\Omega^2}{(\omega_k^2+\Omega^2)}\sum_{\pm}[-\vec{\tilde{\sigma}}_{\pm}(\omega_{-k})\cdot\vec{\tilde{\sigma}}_{\pm}(\omega_{k})\\
   + \vec{\tilde{\sigma}}^T_{\pm}(\omega_{-k}) G_\pm (G_+^\dagger G_+ + G_-^\dagger G_-)^{-1} G^\dagger_\pm\vec{\tilde{\sigma}}_{\pm}(\omega_{k}) \\
   + \vec{\tilde{\sigma}}^T_{\mp}(\omega_{-k}) G_\mp (G_+^\dagger G_+ + G_-^\dagger G_-)^{-1} G^\dagger_\pm\vec{\tilde{\sigma}}_{\pm}(\omega_{k}) ]\}.
   \end{split}
\end{equation}
Realizing that $Mu_0^2\Omega^2/2=E_J(\alpha)$ and going back from the vector notation to indexed notation we arrive at Eq.(\ref{effective-spin1}) of the main text.

\section{The Ising spin-spin interaction in the infinite Kagome lattice}
For spatially infinite  Kagome lattice, one can diagonalize the constraint dependent interaction matrices $G$ by the Fourier transformation on the lattice.
\begin{equation}
\tilde{\varphi}_{lm\pm}(\omega_k)=\dfrac{1}{(2\pi)^2}\int_{q = -\pi}^\pi\int_{ r = -\pi}^\pi dq dr z_{qr\pm}(\omega_k)e^{i(lq+mr)}  \end{equation}
\begin{equation}
\tilde{p}_{lm}(\omega_k)=\dfrac{1}{(2\pi)^2}\int_{q = -\pi}^\pi\int_{ r = -\pi}^\pi dq dr {\pi}_{qr}(\omega_k)e^{i(lq+mr)}
\end{equation}
\begin{equation}
\tilde{\sigma}_{lm\pm}(\omega_k)=\dfrac{1}{(2\pi)^2}\int_{q = -\pi}^\pi \int_{r = -\pi}^\pi dq dr {\Sigma}_{qr\pm}(\omega_k)e^{i(lq+mr)}
\end{equation}
With these constraints defined in Eq.(\ref{G-function}) we obtain expressions for the entries of the diagonalized constraint matrices in the Fourier space
\begin{equation}
\begin{split}
    \mathcal{G}_{qr+} = 1+e^{ir}-2e^{iq}\\
    \mathcal{G}_{qr-}= e^{i(r+q)}+e^{iq}-2e^{ir}.
\end{split}
\end{equation}
and the modulus of the entries does not depend on the sign index $\pm$.
\begin{equation}
\begin{split}
    |\mathcal{G}_{qr}|^2 = \mathcal{G}_{qr\pm}^*\mathcal{G}_{qr\pm}= \\
    -4\sin^2(r/2)+8\sin^2((q-r)/2)+8\sin^2((q)/2).
\end{split}
\end{equation}
Since the matrices are diagonal we can easily calculate their inverse and express the interaction matrices (\ref{G-pm}) and (\ref{G-mp}) as
\begin{equation}\label{Intra}
\begin{split}
    G_{lml'm'}^{\pm\pm}= \dfrac{1}{(2\pi)^2}\iint_{-\pi}^\pi \dfrac{\mathcal{G}_{qr\pm} \mathcal{G}_{qr\pm}^*}{2|\mathcal{G}_{qr}|^2 }e^{-i[q(l-l')+r(m-m')]}drdq \\
    =\dfrac{1}{(2\pi)^2}\iint_{-\pi}^\pi\dfrac{1}{2}e^{-i[q(l-l')+r(m-m')]}drdq =\dfrac{1}{2}\delta_{ll'}\delta_{mm'}.
\end{split}
\end{equation}
and
\begin{equation}\label{Inter}
\begin{split}
    G_{lml'm'}^{\pm\mp}=G_{l'm'lm}^{\mp\pm}\\
    \dfrac{1}{2(2\pi)^2}\iint_{-\pi}^\pi \dfrac{\mathcal{G}_{qr\pm} \mathcal{G}_{qr\mp}^*}{2|\mathcal{G}_{qr}|^2 }e^{-i[q(l-l')+r(m-m')]}drdq\\
    =\dfrac{1}{(2\pi)^2}\iint_{-\pi}^\pi\dfrac{1+e^{ir}-2e^{iq}}{e^{i(q+r)}+e^{iq}-2e^{ir}}e^{-i[q(l-l')+r(m-m')]}drdq.
\end{split}
\end{equation}
As one can see from Eq.(\ref{Intra}) the interaction between different spins within one sublattice vanishes in the infinite system. We are thus left with interactions between the sublattices given by Eq.(\ref{Inter}). To evaluate the integral we perform the substitution $z_r=e^{ir}$ and $z_q=e^{iq}$ and obtain
\begin{equation}\label{InterZ}
\begin{split}
    2(2\pi)^2 G_{lml'm'}^{\pm\mp} = 2(2\pi)^2 G_{l'm'lm}^{\mp\pm}\\
    =-\oint_{S_1}\oint_{S_1} \dfrac{z^{-1}_r z^{-1}_q+z^{-1}_q-2z^{-1}_r}{z_r z_q +z_q-2z_r}z_q^{-(l-l')}z_r^{-(m-m')}dz_rdz_q \\
    =-\oint_{S_1}\oint_{S_1}  \dfrac{1}{z_r z_q +z_q-2z_r}z_q^{-(l-l'+1)}z_r^{-(m-m'+1)}dz_rdz_q \\
    -\oint_{S_1}\oint_{S_1} \dfrac{1}{z_r z_q +z_q-2z_r}z_q^{-(l-l'+1)}z_r^{-(m-m')}dz_rdz_q \\
    +\oint_{S_1}\oint_{S_1} \dfrac{2}{z_r z_q +z_q-2_zr}z_q^{-(l-l')}z_r^{-(m-m'+1)}dz_rdz_q \\
    = [1] + [2] + [3],
\end{split}
\end{equation}
where we labeled the last three integrals as $[1],[2],[3]$. First we perform the integration over $z_q$ and then over $z_r$. By virtue of the residue theorem we can compute the integral by evaluating the residue of the poles within the unit circle. We start by investigating the function $f_n(z_q)$
\begin{equation}
f_n(z_q)=\dfrac{1}{z_r z_q+z_q -2z_r}z_q^{-n}
\end{equation}
There are only two poles for $z_q$. The poles at $z_{q0}=0$ and $z_{q1}=2z_r/(1+z_r)$. Since $|z_q|>1$ for almost all $z_r\in S_1$, its contribution will vanish when performing the integral over $z_r$. Thus it is sufficient to calculate the residue at $z_{q0}$ only
\begin{equation}
\begin{split}
\text{Res}(f_n,0)=\dfrac{1}{(n-1)!}\lim_{z_q \rightarrow 0}\dfrac{d^{n-1}}{dz_q^{n-1}}\dfrac{1}{z_r z_q+z_q -2z_r}  \\
=\dfrac{1}{(n-1)!}\lim_{z_q \rightarrow 0}\dfrac{(n-1)!(-1)^{n-1}(z_r+1)^{n-1}}{(z_r z_q+z_q -2z_r)^n}\\
=\dfrac{(-1)^{n-1}(z_r+1)^{n-1}}{(-2z_r)^n}=\dfrac{-(z_r+1)^{n-1}}{2^n z_r^n}.
\end{split}
\end{equation}
To perform integration over $z_r$, we will take a look at the function $g_{nh}(z_r)$. 
\begin{equation}
g_{nh}(z_r)=(z_r+1)^{h} z_r^{-n}.
\end{equation}
In case of $h\geq0$ we are only left with the residue at $z_{r0}=0$
\begin{equation}
\begin{split}
\text{Res}(g,0)=\dfrac{1}{(n-1)!}\lim_{z_r \rightarrow 0}\dfrac{d^{n-1}}{dz_q^{n-1}}(z_r+1)^{h}\\
=\dfrac{1}{(n-1)!}\lim_{z_r \rightarrow 0}(z_r+1)^{h-(n-1)}\dfrac{h!}{(h-(n-1))!}\\
=\binom{h}{n-1}\large|_{h \geq n-1}.
\end{split}
\end{equation}
With this in mind we can finally evaluate our inter lattice interaction
\begin{equation}\label{InterF}
\begin{split}
    G_{lml'm'}^{\pm\mp}=G_{l'm'lm}^{\mp\pm}\\
    =-\dfrac{1}{2}\dfrac{1}{2^{l-l'}}\binom{l-l'}{l-l'+m-m'+1}  \large|_{l-l'\geq 0};_{-1\geq m-m'\geq l'-l-1}\\
    -\dfrac{1}{2}\dfrac{1}{2^{l-l'}}\binom{l-l'}{m-m'+l-l'}\large|_{l-l'\geq 0};_{0 \geq m-m'\geq l'-l}\\
    +2\dfrac{1}{2^{l-l'}}\binom{l-l'-1}{m-m'+l-l'}\large|_{l-l'\geq 1;-1 \geq m-m'\geq l'-l }
\end{split}
\end{equation}

\bibliography{apssamp,general,
flatband,josephson}

\providecommand{\noopsort}[1]{}\providecommand{\singleletter}[1]{#1}%
\begin{thebibliography}{50}%
\makeatletter
\providecommand \@ifxundefined [1]{%
 \@ifx{#1\undefined}
}%
\providecommand \@ifnum [1]{%
 \ifnum #1\expandafter \@firstoftwo
 \else \expandafter \@secondoftwo
 \fi
}%
\providecommand \@ifx [1]{%
 \ifx #1\expandafter \@firstoftwo
 \else \expandafter \@secondoftwo
 \fi
}%
\providecommand \natexlab [1]{#1}%
\providecommand \enquote  [1]{``#1''}%
\providecommand \bibnamefont  [1]{#1}%
\providecommand \bibfnamefont [1]{#1}%
\providecommand \citenamefont [1]{#1}%
\providecommand \href@noop [0]{\@secondoftwo}%
\providecommand \href [0]{\begingroup \@sanitize@url \@href}%
\providecommand \@href[1]{\@@startlink{#1}\@@href}%
\providecommand \@@href[1]{\endgroup#1\@@endlink}%
\providecommand \@sanitize@url [0]{\catcode `\\12\catcode `\$12\catcode `\&12\catcode `\#12\catcode `\^12\catcode `\_12\catcode `\%12\relax}%
\providecommand \@@startlink[1]{}%
\providecommand \@@endlink[0]{}%
\providecommand \url  [0]{\begingroup\@sanitize@url \@url }%
\providecommand \@url [1]{\endgroup\@href {#1}{\urlprefix }}%
\providecommand \urlprefix  [0]{URL }%
\providecommand \Eprint [0]{\href }%
\providecommand \doibase [0]{https://doi.org/}%
\providecommand \selectlanguage [0]{\@gobble}%
\providecommand \bibinfo  [0]{\@secondoftwo}%
\providecommand \bibfield  [0]{\@secondoftwo}%
\providecommand \translation [1]{[#1]}%
\providecommand \BibitemOpen [0]{}%
\providecommand \bibitemStop [0]{}%
\providecommand \bibitemNoStop [0]{.\EOS\space}%
\providecommand \EOS [0]{\spacefactor3000\relax}%
\providecommand \BibitemShut  [1]{\csname bibitem#1\endcsname}%
\let\auto@bib@innerbib\@empty
\bibitem [{\citenamefont {Anderson}(1978)}]{anderson1978concept}%
  \BibitemOpen
  \bibfield  {author} {\bibinfo {author} {\bibfnamefont {P.}~\bibnamefont {Anderson}},\ }\bibfield  {title} {\bibinfo {title} {The concept of frustration in spin glasses},\ }\href@noop {} {\bibfield  {journal} {\bibinfo  {journal} {Journal of the Less Common Metals}\ }\textbf {\bibinfo {volume} {62}},\ \bibinfo {pages} {291} (\bibinfo {year} {1978})}\BibitemShut {NoStop}%
\bibitem [{\citenamefont {Moessner}\ and\ \citenamefont {Ramirez}(2006)}]{moessner2006geometrical}%
  \BibitemOpen
  \bibfield  {author} {\bibinfo {author} {\bibfnamefont {R.}~\bibnamefont {Moessner}}\ and\ \bibinfo {author} {\bibfnamefont {A.~P.}\ \bibnamefont {Ramirez}},\ }\bibfield  {title} {\bibinfo {title} {Geometrical frustration},\ }\href@noop {} {\bibfield  {journal} {\bibinfo  {journal} {Phys. Today}\ }\textbf {\bibinfo {volume} {59}},\ \bibinfo {pages} {24} (\bibinfo {year} {2006})}\BibitemShut {NoStop}%
\bibitem [{\citenamefont {Schr{\"o}der}\ \emph {et~al.}(2005)\citenamefont {Schr{\"o}der}, \citenamefont {Nojiri}, \citenamefont {Schnack}, \citenamefont {Hage}, \citenamefont {Luban},\ and\ \citenamefont {K{\"o}gerler}}]{schroder2005competing}%
  \BibitemOpen
  \bibfield  {author} {\bibinfo {author} {\bibfnamefont {C.}~\bibnamefont {Schr{\"o}der}}, \bibinfo {author} {\bibfnamefont {H.}~\bibnamefont {Nojiri}}, \bibinfo {author} {\bibfnamefont {J.}~\bibnamefont {Schnack}}, \bibinfo {author} {\bibfnamefont {P.}~\bibnamefont {Hage}}, \bibinfo {author} {\bibfnamefont {M.}~\bibnamefont {Luban}},\ and\ \bibinfo {author} {\bibfnamefont {P.}~\bibnamefont {K{\"o}gerler}},\ }\bibfield  {title} {\bibinfo {title} {Competing spin phases in geometrically frustrated magnetic molecules},\ }\href@noop {} {\bibfield  {journal} {\bibinfo  {journal} {Physical Review Letters}\ }\textbf {\bibinfo {volume} {94}},\ \bibinfo {pages} {017205} (\bibinfo {year} {2005})}\BibitemShut {NoStop}%
\bibitem [{\citenamefont {Balents}(2010)}]{balents2010spin}%
  \BibitemOpen
  \bibfield  {author} {\bibinfo {author} {\bibfnamefont {L.}~\bibnamefont {Balents}},\ }\bibfield  {title} {\bibinfo {title} {Spin liquids in frustrated magnets},\ }\href@noop {} {\bibfield  {journal} {\bibinfo  {journal} {Nature}\ }\textbf {\bibinfo {volume} {464}},\ \bibinfo {pages} {199} (\bibinfo {year} {2010})}\BibitemShut {NoStop}%
\bibitem [{\citenamefont {Baniodeh}\ \emph {et~al.}(2018)\citenamefont {Baniodeh}, \citenamefont {Magnani}, \citenamefont {Lan}, \citenamefont {Buth}, \citenamefont {Anson}, \citenamefont {Richter}, \citenamefont {Affronte}, \citenamefont {Schnack},\ and\ \citenamefont {Powell}}]{baniodeh2018high}%
  \BibitemOpen
  \bibfield  {author} {\bibinfo {author} {\bibfnamefont {A.}~\bibnamefont {Baniodeh}}, \bibinfo {author} {\bibfnamefont {N.}~\bibnamefont {Magnani}}, \bibinfo {author} {\bibfnamefont {Y.}~\bibnamefont {Lan}}, \bibinfo {author} {\bibfnamefont {G.}~\bibnamefont {Buth}}, \bibinfo {author} {\bibfnamefont {C.~E.}\ \bibnamefont {Anson}}, \bibinfo {author} {\bibfnamefont {J.}~\bibnamefont {Richter}}, \bibinfo {author} {\bibfnamefont {M.}~\bibnamefont {Affronte}}, \bibinfo {author} {\bibfnamefont {J.}~\bibnamefont {Schnack}},\ and\ \bibinfo {author} {\bibfnamefont {A.~K.}\ \bibnamefont {Powell}},\ }\bibfield  {title} {\bibinfo {title} {High spin cycles: topping the spin record for a single molecule verging on quantum criticality},\ }\href@noop {} {\bibfield  {journal} {\bibinfo  {journal} {npj Quantum Materials}\ }\textbf {\bibinfo {volume} {3}},\ \bibinfo {pages} {10} (\bibinfo {year} {2018})}\BibitemShut {NoStop}%
\bibitem [{\citenamefont {Han}\ \emph {et~al.}(2012)\citenamefont {Han}, \citenamefont {Helton}, \citenamefont {Chu}, \citenamefont {Nocera}, \citenamefont {Rodriguez-Rivera}, \citenamefont {Broholm},\ and\ \citenamefont {Lee}}]{han2012fractionalized}%
  \BibitemOpen
  \bibfield  {author} {\bibinfo {author} {\bibfnamefont {T.-H.}\ \bibnamefont {Han}}, \bibinfo {author} {\bibfnamefont {J.~S.}\ \bibnamefont {Helton}}, \bibinfo {author} {\bibfnamefont {S.}~\bibnamefont {Chu}}, \bibinfo {author} {\bibfnamefont {D.~G.}\ \bibnamefont {Nocera}}, \bibinfo {author} {\bibfnamefont {J.~A.}\ \bibnamefont {Rodriguez-Rivera}}, \bibinfo {author} {\bibfnamefont {C.}~\bibnamefont {Broholm}},\ and\ \bibinfo {author} {\bibfnamefont {Y.~S.}\ \bibnamefont {Lee}},\ }\bibfield  {title} {\bibinfo {title} {Fractionalized excitations in the spin-liquid state of a kagome-lattice antiferromagnet},\ }\href@noop {} {\bibfield  {journal} {\bibinfo  {journal} {Nature}\ }\textbf {\bibinfo {volume} {492}},\ \bibinfo {pages} {406} (\bibinfo {year} {2012})}\BibitemShut {NoStop}%
\bibitem [{\citenamefont {Song}\ and\ \citenamefont {Zhang}(2023)}]{song2023tensor}%
  \BibitemOpen
  \bibfield  {author} {\bibinfo {author} {\bibfnamefont {F.-F.}\ \bibnamefont {Song}}\ and\ \bibinfo {author} {\bibfnamefont {G.-M.}\ \bibnamefont {Zhang}},\ }\href@noop {} {\bibinfo {title} {Tensor network approach to the fully frustrated xy model on a kagome lattice with a fractional vortex-antivortex pairing transition}} (\bibinfo {year} {2023}),\ \Eprint {https://arxiv.org/abs/2304.11944} {arXiv:2304.11944 [cond-mat.str-el]} \BibitemShut {NoStop}%
\bibitem [{\citenamefont {Nisoli}\ \emph {et~al.}(2013)\citenamefont {Nisoli}, \citenamefont {Moessner},\ and\ \citenamefont {Schiffer}}]{nisoli2013colloquium}%
  \BibitemOpen
  \bibfield  {author} {\bibinfo {author} {\bibfnamefont {C.}~\bibnamefont {Nisoli}}, \bibinfo {author} {\bibfnamefont {R.}~\bibnamefont {Moessner}},\ and\ \bibinfo {author} {\bibfnamefont {P.}~\bibnamefont {Schiffer}},\ }\bibfield  {title} {\bibinfo {title} {Colloquium: Artificial spin ice: Designing and imaging magnetic frustration},\ }\href@noop {} {\bibfield  {journal} {\bibinfo  {journal} {Reviews of Modern Physics}\ }\textbf {\bibinfo {volume} {85}},\ \bibinfo {pages} {1473} (\bibinfo {year} {2013})}\BibitemShut {NoStop}%
\bibitem [{\citenamefont {Mahmoudian}\ \emph {et~al.}(2015)\citenamefont {Mahmoudian}, \citenamefont {Rademaker}, \citenamefont {Ralko}, \citenamefont {Fratini},\ and\ \citenamefont {Dobrosavljevi{\'c}}}]{mahmoudian2015glassy}%
  \BibitemOpen
  \bibfield  {author} {\bibinfo {author} {\bibfnamefont {S.}~\bibnamefont {Mahmoudian}}, \bibinfo {author} {\bibfnamefont {L.}~\bibnamefont {Rademaker}}, \bibinfo {author} {\bibfnamefont {A.}~\bibnamefont {Ralko}}, \bibinfo {author} {\bibfnamefont {S.}~\bibnamefont {Fratini}},\ and\ \bibinfo {author} {\bibfnamefont {V.}~\bibnamefont {Dobrosavljevi{\'c}}},\ }\bibfield  {title} {\bibinfo {title} {Glassy dynamics in geometrically frustrated coulomb liquids without disorder},\ }\href@noop {} {\bibfield  {journal} {\bibinfo  {journal} {Physical Review Letters}\ }\textbf {\bibinfo {volume} {115}},\ \bibinfo {pages} {025701} (\bibinfo {year} {2015})}\BibitemShut {NoStop}%
\bibitem [{\citenamefont {Gong}\ \emph {et~al.}(2017)\citenamefont {Gong}, \citenamefont {Zhu}, \citenamefont {Sheng},\ and\ \citenamefont {Yang}}]{gong2017possible}%
  \BibitemOpen
  \bibfield  {author} {\bibinfo {author} {\bibfnamefont {S.-S.}\ \bibnamefont {Gong}}, \bibinfo {author} {\bibfnamefont {W.}~\bibnamefont {Zhu}}, \bibinfo {author} {\bibfnamefont {D.}~\bibnamefont {Sheng}},\ and\ \bibinfo {author} {\bibfnamefont {K.}~\bibnamefont {Yang}},\ }\bibfield  {title} {\bibinfo {title} {Possible nematic spin liquid in spin-1 antiferromagnetic system on the square lattice: Implications for the nematic paramagnetic state of fese},\ }\href@noop {} {\bibfield  {journal} {\bibinfo  {journal} {Physical Review B}\ }\textbf {\bibinfo {volume} {95}},\ \bibinfo {pages} {205132} (\bibinfo {year} {2017})}\BibitemShut {NoStop}%
\bibitem [{\citenamefont {Fernandes}\ \emph {et~al.}(2014)\citenamefont {Fernandes}, \citenamefont {Chubukov},\ and\ \citenamefont {Schmalian}}]{fernandes2014drives}%
  \BibitemOpen
  \bibfield  {author} {\bibinfo {author} {\bibfnamefont {R.}~\bibnamefont {Fernandes}}, \bibinfo {author} {\bibfnamefont {A.}~\bibnamefont {Chubukov}},\ and\ \bibinfo {author} {\bibfnamefont {J.}~\bibnamefont {Schmalian}},\ }\bibfield  {title} {\bibinfo {title} {What drives nematic order in iron-based superconductors?},\ }\href@noop {} {\bibfield  {journal} {\bibinfo  {journal} {Nature physics}\ }\textbf {\bibinfo {volume} {10}},\ \bibinfo {pages} {97} (\bibinfo {year} {2014})}\BibitemShut {NoStop}%
\bibitem [{\citenamefont {Zhitomirsky}\ and\ \citenamefont {Tsunetsugu}(2010)}]{zhitomirsky2010magnon}%
  \BibitemOpen
  \bibfield  {author} {\bibinfo {author} {\bibfnamefont {M.}~\bibnamefont {Zhitomirsky}}\ and\ \bibinfo {author} {\bibfnamefont {H.}~\bibnamefont {Tsunetsugu}},\ }\bibfield  {title} {\bibinfo {title} {Magnon pairing in quantum spin nematic},\ }\href@noop {} {\bibfield  {journal} {\bibinfo  {journal} {Europhysics Letters}\ }\textbf {\bibinfo {volume} {92}},\ \bibinfo {pages} {37001} (\bibinfo {year} {2010})}\BibitemShut {NoStop}%
\bibitem [{\citenamefont {Yan}\ \emph {et~al.}(2011)\citenamefont {Yan}, \citenamefont {Huse},\ and\ \citenamefont {White}}]{Yan2011}%
  \BibitemOpen
  \bibfield  {author} {\bibinfo {author} {\bibfnamefont {S.}~\bibnamefont {Yan}}, \bibinfo {author} {\bibfnamefont {D.~A.}\ \bibnamefont {Huse}},\ and\ \bibinfo {author} {\bibfnamefont {S.~R.}\ \bibnamefont {White}},\ }\bibfield  {title} {\bibinfo {title} {Spin-liquid ground state of the $s= 1/2$ kagome heisenberg antiferromagnet},\ }\href {https://doi.org/10.1126/science.1201080} {\bibfield  {journal} {\bibinfo  {journal} {Science}\ }\textbf {\bibinfo {volume} {332}},\ \bibinfo {pages} {1173} (\bibinfo {year} {2011})}\BibitemShut {NoStop}%
\bibitem [{\citenamefont {Yamada}\ \emph {et~al.}(2016)\citenamefont {Yamada}, \citenamefont {Soejima}, \citenamefont {Tsuji}, \citenamefont {Hirai}, \citenamefont {Dinc{\u{a}}},\ and\ \citenamefont {Aoki}}]{yamada2016first}%
  \BibitemOpen
  \bibfield  {author} {\bibinfo {author} {\bibfnamefont {M.~G.}\ \bibnamefont {Yamada}}, \bibinfo {author} {\bibfnamefont {T.}~\bibnamefont {Soejima}}, \bibinfo {author} {\bibfnamefont {N.}~\bibnamefont {Tsuji}}, \bibinfo {author} {\bibfnamefont {D.}~\bibnamefont {Hirai}}, \bibinfo {author} {\bibfnamefont {M.}~\bibnamefont {Dinc{\u{a}}}},\ and\ \bibinfo {author} {\bibfnamefont {H.}~\bibnamefont {Aoki}},\ }\bibfield  {title} {\bibinfo {title} {First-principles design of a half-filled flat band of the kagome lattice in two-dimensional metal-organic frameworks},\ }\href@noop {} {\bibfield  {journal} {\bibinfo  {journal} {Physical Review B}\ }\textbf {\bibinfo {volume} {94}},\ \bibinfo {pages} {081102} (\bibinfo {year} {2016})}\BibitemShut {NoStop}%
\bibitem [{\citenamefont {Messio}\ \emph {et~al.}(2012)\citenamefont {Messio}, \citenamefont {Bernu},\ and\ \citenamefont {Lhuillier}}]{messio2012kagome}%
  \BibitemOpen
  \bibfield  {author} {\bibinfo {author} {\bibfnamefont {L.}~\bibnamefont {Messio}}, \bibinfo {author} {\bibfnamefont {B.}~\bibnamefont {Bernu}},\ and\ \bibinfo {author} {\bibfnamefont {C.}~\bibnamefont {Lhuillier}},\ }\bibfield  {title} {\bibinfo {title} {Kagome antiferromagnet: a chiral topological spin liquid?},\ }\href@noop {} {\bibfield  {journal} {\bibinfo  {journal} {Physical Review Letters}\ }\textbf {\bibinfo {volume} {108}},\ \bibinfo {pages} {207204} (\bibinfo {year} {2012})}\BibitemShut {NoStop}%
\bibitem [{\citenamefont {Fujihala}\ \emph {et~al.}(2020)\citenamefont {Fujihala}, \citenamefont {Morita}, \citenamefont {Mole}, \citenamefont {Mitsuda}, \citenamefont {Tohyama}, \citenamefont {ichiro Yano}, \citenamefont {Yu}, \citenamefont {Sota}, \citenamefont {Kuwai}, \citenamefont {Koda}, \citenamefont {Okabe}, \citenamefont {Lee}, \citenamefont {Itoh}, \citenamefont {Hawai}, \citenamefont {Masuda}, \citenamefont {Sagayama}, \citenamefont {Matsuo}, \citenamefont {Kindo}, \citenamefont {Ohira-Kawamura},\ and\ \citenamefont {Nakajima}}]{Fujihala2020}%
  \BibitemOpen
  \bibfield  {author} {\bibinfo {author} {\bibfnamefont {M.}~\bibnamefont {Fujihala}}, \bibinfo {author} {\bibfnamefont {K.}~\bibnamefont {Morita}}, \bibinfo {author} {\bibfnamefont {R.}~\bibnamefont {Mole}}, \bibinfo {author} {\bibfnamefont {S.}~\bibnamefont {Mitsuda}}, \bibinfo {author} {\bibfnamefont {T.}~\bibnamefont {Tohyama}}, \bibinfo {author} {\bibfnamefont {S.}~\bibnamefont {ichiro Yano}}, \bibinfo {author} {\bibfnamefont {D.}~\bibnamefont {Yu}}, \bibinfo {author} {\bibfnamefont {S.}~\bibnamefont {Sota}}, \bibinfo {author} {\bibfnamefont {T.}~\bibnamefont {Kuwai}}, \bibinfo {author} {\bibfnamefont {A.}~\bibnamefont {Koda}}, \bibinfo {author} {\bibfnamefont {H.}~\bibnamefont {Okabe}}, \bibinfo {author} {\bibfnamefont {H.}~\bibnamefont {Lee}}, \bibinfo {author} {\bibfnamefont {S.}~\bibnamefont {Itoh}}, \bibinfo {author} {\bibfnamefont {T.}~\bibnamefont {Hawai}}, \bibinfo {author} {\bibfnamefont {T.}~\bibnamefont {Masuda}}, \bibinfo {author} {\bibfnamefont {H.}~\bibnamefont {Sagayama}}, \bibinfo
  {author} {\bibfnamefont {A.}~\bibnamefont {Matsuo}}, \bibinfo {author} {\bibfnamefont {K.}~\bibnamefont {Kindo}}, \bibinfo {author} {\bibfnamefont {S.}~\bibnamefont {Ohira-Kawamura}},\ and\ \bibinfo {author} {\bibfnamefont {K.}~\bibnamefont {Nakajima}},\ }\bibfield  {title} {\bibinfo {title} {Gapless spin liquid in a square-kagome lattice antiferromagnet},\ }\bibfield  {journal} {\bibinfo  {journal} {Nature Communications}\ }\textbf {\bibinfo {volume} {11}},\ \href {https://doi.org/10.1038/s41467-020-17235-z} {10.1038/s41467-020-17235-z} (\bibinfo {year} {2020})\BibitemShut {NoStop}%
\bibitem [{\citenamefont {Teng}\ \emph {et~al.}(2023)\citenamefont {Teng}, \citenamefont {Oh}, \citenamefont {Tan}, \citenamefont {Chen}, \citenamefont {Huang}, \citenamefont {Gao}, \citenamefont {Yin}, \citenamefont {Chu}, \citenamefont {Hashimoto}, \citenamefont {Lu}, \citenamefont {Jozwiak}, \citenamefont {Bostwick}, \citenamefont {Rotenberg}, \citenamefont {Granroth}, \citenamefont {Yan}, \citenamefont {Birgeneau}, \citenamefont {Dai},\ and\ \citenamefont {Yi}}]{Teng2023}%
  \BibitemOpen
  \bibfield  {author} {\bibinfo {author} {\bibfnamefont {X.}~\bibnamefont {Teng}}, \bibinfo {author} {\bibfnamefont {J.~S.}\ \bibnamefont {Oh}}, \bibinfo {author} {\bibfnamefont {H.}~\bibnamefont {Tan}}, \bibinfo {author} {\bibfnamefont {L.}~\bibnamefont {Chen}}, \bibinfo {author} {\bibfnamefont {J.}~\bibnamefont {Huang}}, \bibinfo {author} {\bibfnamefont {B.}~\bibnamefont {Gao}}, \bibinfo {author} {\bibfnamefont {J.-X.}\ \bibnamefont {Yin}}, \bibinfo {author} {\bibfnamefont {J.-H.}\ \bibnamefont {Chu}}, \bibinfo {author} {\bibfnamefont {M.}~\bibnamefont {Hashimoto}}, \bibinfo {author} {\bibfnamefont {D.}~\bibnamefont {Lu}}, \bibinfo {author} {\bibfnamefont {C.}~\bibnamefont {Jozwiak}}, \bibinfo {author} {\bibfnamefont {A.}~\bibnamefont {Bostwick}}, \bibinfo {author} {\bibfnamefont {E.}~\bibnamefont {Rotenberg}}, \bibinfo {author} {\bibfnamefont {G.~E.}\ \bibnamefont {Granroth}}, \bibinfo {author} {\bibfnamefont {B.}~\bibnamefont {Yan}}, \bibinfo {author} {\bibfnamefont {R.~J.}\ \bibnamefont {Birgeneau}},
  \bibinfo {author} {\bibfnamefont {P.}~\bibnamefont {Dai}},\ and\ \bibinfo {author} {\bibfnamefont {M.}~\bibnamefont {Yi}},\ }\bibfield  {title} {\bibinfo {title} {Magnetism and charge density wave order in kagome {FeGe}},\ }\href {https://doi.org/10.1038/s41567-023-01985-w} {\bibfield  {journal} {\bibinfo  {journal} {Nature Physics}\ }\textbf {\bibinfo {volume} {19}},\ \bibinfo {pages} {814} (\bibinfo {year} {2023})}\BibitemShut {NoStop}%
\bibitem [{\citenamefont {Neupert}\ \emph {et~al.}(2022)\citenamefont {Neupert}, \citenamefont {Denner}, \citenamefont {Yin}, \citenamefont {Thomale},\ and\ \citenamefont {Hasan}}]{neupert2022charge}%
  \BibitemOpen
  \bibfield  {author} {\bibinfo {author} {\bibfnamefont {T.}~\bibnamefont {Neupert}}, \bibinfo {author} {\bibfnamefont {M.~M.}\ \bibnamefont {Denner}}, \bibinfo {author} {\bibfnamefont {J.-X.}\ \bibnamefont {Yin}}, \bibinfo {author} {\bibfnamefont {R.}~\bibnamefont {Thomale}},\ and\ \bibinfo {author} {\bibfnamefont {M.~Z.}\ \bibnamefont {Hasan}},\ }\bibfield  {title} {\bibinfo {title} {Charge order and superconductivity in kagome materials},\ }\href@noop {} {\bibfield  {journal} {\bibinfo  {journal} {Nature Physics}\ }\textbf {\bibinfo {volume} {18}},\ \bibinfo {pages} {137} (\bibinfo {year} {2022})}\BibitemShut {NoStop}%
\bibitem [{\citenamefont {Feng}\ \emph {et~al.}(2021)\citenamefont {Feng}, \citenamefont {Jiang}, \citenamefont {Wang},\ and\ \citenamefont {Hu}}]{feng2021chiral}%
  \BibitemOpen
  \bibfield  {author} {\bibinfo {author} {\bibfnamefont {X.}~\bibnamefont {Feng}}, \bibinfo {author} {\bibfnamefont {K.}~\bibnamefont {Jiang}}, \bibinfo {author} {\bibfnamefont {Z.}~\bibnamefont {Wang}},\ and\ \bibinfo {author} {\bibfnamefont {J.}~\bibnamefont {Hu}},\ }\bibfield  {title} {\bibinfo {title} {Chiral flux phase in the kagome superconductor av3sb5},\ }\href@noop {} {\bibfield  {journal} {\bibinfo  {journal} {Science bulletin}\ }\textbf {\bibinfo {volume} {66}},\ \bibinfo {pages} {1384} (\bibinfo {year} {2021})}\BibitemShut {NoStop}%
\bibitem [{\citenamefont {Yin}\ \emph {et~al.}(2022)\citenamefont {Yin}, \citenamefont {Lian},\ and\ \citenamefont {Hasan}}]{yin2022topological}%
  \BibitemOpen
  \bibfield  {author} {\bibinfo {author} {\bibfnamefont {J.-X.}\ \bibnamefont {Yin}}, \bibinfo {author} {\bibfnamefont {B.}~\bibnamefont {Lian}},\ and\ \bibinfo {author} {\bibfnamefont {M.~Z.}\ \bibnamefont {Hasan}},\ }\bibfield  {title} {\bibinfo {title} {Topological kagome magnets and superconductors},\ }\href@noop {} {\bibfield  {journal} {\bibinfo  {journal} {Nature}\ }\textbf {\bibinfo {volume} {612}},\ \bibinfo {pages} {647} (\bibinfo {year} {2022})}\BibitemShut {NoStop}%
\bibitem [{\citenamefont {Yang}\ \emph {et~al.}(2023)\citenamefont {Yang}, \citenamefont {Yi}, \citenamefont {Zhao}, \citenamefont {Xie}, \citenamefont {Miao}, \citenamefont {Luo}, \citenamefont {Chen}, \citenamefont {Liang}, \citenamefont {Zhu}, \citenamefont {Ye}, \citenamefont {You}, \citenamefont {Gu}, \citenamefont {Zhang}, \citenamefont {Zhang}, \citenamefont {Yang}, \citenamefont {Wang}, \citenamefont {Peng}, \citenamefont {Mao}, \citenamefont {Liu}, \citenamefont {Xu}, \citenamefont {Chen}, \citenamefont {Yang}, \citenamefont {Su}, \citenamefont {Gao}, \citenamefont {Zhao},\ and\ \citenamefont {Zhou}}]{Yang2023}%
  \BibitemOpen
  \bibfield  {author} {\bibinfo {author} {\bibfnamefont {J.}~\bibnamefont {Yang}}, \bibinfo {author} {\bibfnamefont {X.}~\bibnamefont {Yi}}, \bibinfo {author} {\bibfnamefont {Z.}~\bibnamefont {Zhao}}, \bibinfo {author} {\bibfnamefont {Y.}~\bibnamefont {Xie}}, \bibinfo {author} {\bibfnamefont {T.}~\bibnamefont {Miao}}, \bibinfo {author} {\bibfnamefont {H.}~\bibnamefont {Luo}}, \bibinfo {author} {\bibfnamefont {H.}~\bibnamefont {Chen}}, \bibinfo {author} {\bibfnamefont {B.}~\bibnamefont {Liang}}, \bibinfo {author} {\bibfnamefont {W.}~\bibnamefont {Zhu}}, \bibinfo {author} {\bibfnamefont {Y.}~\bibnamefont {Ye}}, \bibinfo {author} {\bibfnamefont {J.-Y.}\ \bibnamefont {You}}, \bibinfo {author} {\bibfnamefont {B.}~\bibnamefont {Gu}}, \bibinfo {author} {\bibfnamefont {S.}~\bibnamefont {Zhang}}, \bibinfo {author} {\bibfnamefont {F.}~\bibnamefont {Zhang}}, \bibinfo {author} {\bibfnamefont {F.}~\bibnamefont {Yang}}, \bibinfo {author} {\bibfnamefont {Z.}~\bibnamefont {Wang}}, \bibinfo {author} {\bibfnamefont
  {Q.}~\bibnamefont {Peng}}, \bibinfo {author} {\bibfnamefont {H.}~\bibnamefont {Mao}}, \bibinfo {author} {\bibfnamefont {G.}~\bibnamefont {Liu}}, \bibinfo {author} {\bibfnamefont {Z.}~\bibnamefont {Xu}}, \bibinfo {author} {\bibfnamefont {H.}~\bibnamefont {Chen}}, \bibinfo {author} {\bibfnamefont {H.}~\bibnamefont {Yang}}, \bibinfo {author} {\bibfnamefont {G.}~\bibnamefont {Su}}, \bibinfo {author} {\bibfnamefont {H.}~\bibnamefont {Gao}}, \bibinfo {author} {\bibfnamefont {L.}~\bibnamefont {Zhao}},\ and\ \bibinfo {author} {\bibfnamefont {X.~J.}\ \bibnamefont {Zhou}},\ }\bibfield  {title} {\bibinfo {title} {Observation of flat band, dirac nodal lines and topological surface states in kagome superconductor {CsTi}3bi5},\ }\bibfield  {journal} {\bibinfo  {journal} {Nature Communications}\ }\textbf {\bibinfo {volume} {14}},\ \href {https://doi.org/10.1038/s41467-023-39620-0} {10.1038/s41467-023-39620-0} (\bibinfo {year} {2023})\BibitemShut {NoStop}%
\bibitem [{\citenamefont {Britton}\ \emph {et~al.}(2012)\citenamefont {Britton}, \citenamefont {Sawyer}, \citenamefont {Keith}, \citenamefont {Wang}, \citenamefont {Freericks}, \citenamefont {Uys}, \citenamefont {Biercuk},\ and\ \citenamefont {Bollinger}}]{britton2012engineered}%
  \BibitemOpen
  \bibfield  {author} {\bibinfo {author} {\bibfnamefont {J.~W.}\ \bibnamefont {Britton}}, \bibinfo {author} {\bibfnamefont {B.~C.}\ \bibnamefont {Sawyer}}, \bibinfo {author} {\bibfnamefont {A.~C.}\ \bibnamefont {Keith}}, \bibinfo {author} {\bibfnamefont {C.-C.~J.}\ \bibnamefont {Wang}}, \bibinfo {author} {\bibfnamefont {J.~K.}\ \bibnamefont {Freericks}}, \bibinfo {author} {\bibfnamefont {H.}~\bibnamefont {Uys}}, \bibinfo {author} {\bibfnamefont {M.~J.}\ \bibnamefont {Biercuk}},\ and\ \bibinfo {author} {\bibfnamefont {J.~J.}\ \bibnamefont {Bollinger}},\ }\bibfield  {title} {\bibinfo {title} {Engineered two-dimensional ising interactions in a trapped-ion quantum simulator with hundreds of spins},\ }\href@noop {} {\bibfield  {journal} {\bibinfo  {journal} {Nature}\ }\textbf {\bibinfo {volume} {484}},\ \bibinfo {pages} {489} (\bibinfo {year} {2012})}\BibitemShut {NoStop}%
\bibitem [{\citenamefont {Weimann}\ \emph {et~al.}(2016)\citenamefont {Weimann}, \citenamefont {Morales-Inostroza}, \citenamefont {Real}, \citenamefont {Cantillano}, \citenamefont {Szameit},\ and\ \citenamefont {Vicencio}}]{weimann2016transport}%
  \BibitemOpen
  \bibfield  {author} {\bibinfo {author} {\bibfnamefont {S.}~\bibnamefont {Weimann}}, \bibinfo {author} {\bibfnamefont {L.}~\bibnamefont {Morales-Inostroza}}, \bibinfo {author} {\bibfnamefont {B.}~\bibnamefont {Real}}, \bibinfo {author} {\bibfnamefont {C.}~\bibnamefont {Cantillano}}, \bibinfo {author} {\bibfnamefont {A.}~\bibnamefont {Szameit}},\ and\ \bibinfo {author} {\bibfnamefont {R.~A.}\ \bibnamefont {Vicencio}},\ }\bibfield  {title} {\bibinfo {title} {Transport in sawtooth photonic lattices},\ }\href {https://doi.org/10.1364/OL.41.002414} {\bibfield  {journal} {\bibinfo  {journal} {Opt. Lett.}\ }\textbf {\bibinfo {volume} {41}},\ \bibinfo {pages} {2414} (\bibinfo {year} {2016})}\BibitemShut {NoStop}%
\bibitem [{\citenamefont {Vicencio}\ \emph {et~al.}(2015)\citenamefont {Vicencio}, \citenamefont {Cantillano}, \citenamefont {Morales-Inostroza}, \citenamefont {Real}, \citenamefont {Mej\'{\i}a-Cort\'es}, \citenamefont {Weimann}, \citenamefont {Szameit},\ and\ \citenamefont {Molina}}]{vicencio2015observation}%
  \BibitemOpen
  \bibfield  {author} {\bibinfo {author} {\bibfnamefont {R.~A.}\ \bibnamefont {Vicencio}}, \bibinfo {author} {\bibfnamefont {C.}~\bibnamefont {Cantillano}}, \bibinfo {author} {\bibfnamefont {L.}~\bibnamefont {Morales-Inostroza}}, \bibinfo {author} {\bibfnamefont {B.}~\bibnamefont {Real}}, \bibinfo {author} {\bibfnamefont {C.}~\bibnamefont {Mej\'{\i}a-Cort\'es}}, \bibinfo {author} {\bibfnamefont {S.}~\bibnamefont {Weimann}}, \bibinfo {author} {\bibfnamefont {A.}~\bibnamefont {Szameit}},\ and\ \bibinfo {author} {\bibfnamefont {M.~I.}\ \bibnamefont {Molina}},\ }\bibfield  {title} {\bibinfo {title} {Observation of localized states in lieb photonic lattices},\ }\href {https://doi.org/10.1103/PhysRevLett.114.245503} {\bibfield  {journal} {\bibinfo  {journal} {Phys. Rev. Lett.}\ }\textbf {\bibinfo {volume} {114}},\ \bibinfo {pages} {245503} (\bibinfo {year} {2015})}\BibitemShut {NoStop}%
\bibitem [{\citenamefont {Semeghini}\ \emph {et~al.}(2021)\citenamefont {Semeghini}, \citenamefont {Levine}, \citenamefont {Keesling}, \citenamefont {Ebadi}, \citenamefont {Wang}, \citenamefont {Bluvstein}, \citenamefont {Verresen}, \citenamefont {Pichler}, \citenamefont {Kalinowski}, \citenamefont {Samajdar}, \citenamefont {Omran}, \citenamefont {Sachdev}, \citenamefont {Vishwanath}, \citenamefont {Greiner}, \citenamefont {Vuleti{\'{c}}},\ and\ \citenamefont {Lukin}}]{Semeghini2021}%
  \BibitemOpen
  \bibfield  {author} {\bibinfo {author} {\bibfnamefont {G.}~\bibnamefont {Semeghini}}, \bibinfo {author} {\bibfnamefont {H.}~\bibnamefont {Levine}}, \bibinfo {author} {\bibfnamefont {A.}~\bibnamefont {Keesling}}, \bibinfo {author} {\bibfnamefont {S.}~\bibnamefont {Ebadi}}, \bibinfo {author} {\bibfnamefont {T.~T.}\ \bibnamefont {Wang}}, \bibinfo {author} {\bibfnamefont {D.}~\bibnamefont {Bluvstein}}, \bibinfo {author} {\bibfnamefont {R.}~\bibnamefont {Verresen}}, \bibinfo {author} {\bibfnamefont {H.}~\bibnamefont {Pichler}}, \bibinfo {author} {\bibfnamefont {M.}~\bibnamefont {Kalinowski}}, \bibinfo {author} {\bibfnamefont {R.}~\bibnamefont {Samajdar}}, \bibinfo {author} {\bibfnamefont {A.}~\bibnamefont {Omran}}, \bibinfo {author} {\bibfnamefont {S.}~\bibnamefont {Sachdev}}, \bibinfo {author} {\bibfnamefont {A.}~\bibnamefont {Vishwanath}}, \bibinfo {author} {\bibfnamefont {M.}~\bibnamefont {Greiner}}, \bibinfo {author} {\bibfnamefont {V.}~\bibnamefont {Vuleti{\'{c}}}},\ and\ \bibinfo {author} {\bibfnamefont
  {M.~D.}\ \bibnamefont {Lukin}},\ }\bibfield  {title} {\bibinfo {title} {Probing topological spin liquids on a programmable quantum simulator},\ }\href {https://doi.org/10.1126/science.abi8794} {\bibfield  {journal} {\bibinfo  {journal} {Science}\ }\textbf {\bibinfo {volume} {374}},\ \bibinfo {pages} {1242} (\bibinfo {year} {2021})}\BibitemShut {NoStop}%
\bibitem [{\citenamefont {Giudici}\ \emph {et~al.}(2022)\citenamefont {Giudici}, \citenamefont {Lukin},\ and\ \citenamefont {Pichler}}]{Giudici2022}%
  \BibitemOpen
  \bibfield  {author} {\bibinfo {author} {\bibfnamefont {G.}~\bibnamefont {Giudici}}, \bibinfo {author} {\bibfnamefont {M.~D.}\ \bibnamefont {Lukin}},\ and\ \bibinfo {author} {\bibfnamefont {H.}~\bibnamefont {Pichler}},\ }\bibfield  {title} {\bibinfo {title} {Dynamical preparation of quantum spin liquids in rydberg atom arrays},\ }\bibfield  {journal} {\bibinfo  {journal} {Physical Review Letters}\ }\textbf {\bibinfo {volume} {129}},\ \href {https://doi.org/10.1103/physrevlett.129.090401} {10.1103/physrevlett.129.090401} (\bibinfo {year} {2022})\BibitemShut {NoStop}%
\bibitem [{\citenamefont {Yan}\ \emph {et~al.}(2023)\citenamefont {Yan}, \citenamefont {Wang}, \citenamefont {Samajdar}, \citenamefont {Sachdev},\ and\ \citenamefont {Meng}}]{Yan2023}%
  \BibitemOpen
  \bibfield  {author} {\bibinfo {author} {\bibfnamefont {Z.}~\bibnamefont {Yan}}, \bibinfo {author} {\bibfnamefont {Y.-C.}\ \bibnamefont {Wang}}, \bibinfo {author} {\bibfnamefont {R.}~\bibnamefont {Samajdar}}, \bibinfo {author} {\bibfnamefont {S.}~\bibnamefont {Sachdev}},\ and\ \bibinfo {author} {\bibfnamefont {Z.~Y.}\ \bibnamefont {Meng}},\ }\bibfield  {title} {\bibinfo {title} {Emergent glassy behavior in a kagome rydberg atom array},\ }\bibfield  {journal} {\bibinfo  {journal} {Physical Review Letters}\ }\textbf {\bibinfo {volume} {130}},\ \href {https://doi.org/10.1103/physrevlett.130.206501} {10.1103/physrevlett.130.206501} (\bibinfo {year} {2023})\BibitemShut {NoStop}%
\bibitem [{\citenamefont {Xu}\ \emph {et~al.}(2023)\citenamefont {Xu}, \citenamefont {Kendrick}, \citenamefont {Kale}, \citenamefont {Gang}, \citenamefont {Ji}, \citenamefont {Scalettar}, \citenamefont {Lebrat},\ and\ \citenamefont {Greiner}}]{Xu2023}%
  \BibitemOpen
  \bibfield  {author} {\bibinfo {author} {\bibfnamefont {M.}~\bibnamefont {Xu}}, \bibinfo {author} {\bibfnamefont {L.~H.}\ \bibnamefont {Kendrick}}, \bibinfo {author} {\bibfnamefont {A.}~\bibnamefont {Kale}}, \bibinfo {author} {\bibfnamefont {Y.}~\bibnamefont {Gang}}, \bibinfo {author} {\bibfnamefont {G.}~\bibnamefont {Ji}}, \bibinfo {author} {\bibfnamefont {R.~T.}\ \bibnamefont {Scalettar}}, \bibinfo {author} {\bibfnamefont {M.}~\bibnamefont {Lebrat}},\ and\ \bibinfo {author} {\bibfnamefont {M.}~\bibnamefont {Greiner}},\ }\bibfield  {title} {\bibinfo {title} {Frustration- and doping-induced magnetism in a fermi{\textendash}hubbard simulator},\ }\bibfield  {journal} {\bibinfo  {journal} {Nature}\ }\href {https://doi.org/10.1038/s41586-023-06280-5} {10.1038/s41586-023-06280-5} (\bibinfo {year} {2023})\BibitemShut {NoStop}%
\bibitem [{\citenamefont {Rzchowski}(1997)}]{rzchowski1997phase}%
  \BibitemOpen
  \bibfield  {author} {\bibinfo {author} {\bibfnamefont {M.~S.}\ \bibnamefont {Rzchowski}},\ }\bibfield  {title} {\bibinfo {title} {Phase transitions in a kagom\'e lattice of josephson junctions},\ }\href {https://doi.org/10.1103/PhysRevB.55.11745} {\bibfield  {journal} {\bibinfo  {journal} {Phys. Rev. B}\ }\textbf {\bibinfo {volume} {55}},\ \bibinfo {pages} {11745} (\bibinfo {year} {1997})}\BibitemShut {NoStop}%
\bibitem [{\citenamefont {Pop}\ \emph {et~al.}(2008)\citenamefont {Pop}, \citenamefont {Hasselbach}, \citenamefont {Buisson}, \citenamefont {Guichard}, \citenamefont {Pannetier},\ and\ \citenamefont {Protopopov}}]{pop2008measurement}%
  \BibitemOpen
  \bibfield  {author} {\bibinfo {author} {\bibfnamefont {I.~M.}\ \bibnamefont {Pop}}, \bibinfo {author} {\bibfnamefont {K.}~\bibnamefont {Hasselbach}}, \bibinfo {author} {\bibfnamefont {O.}~\bibnamefont {Buisson}}, \bibinfo {author} {\bibfnamefont {W.}~\bibnamefont {Guichard}}, \bibinfo {author} {\bibfnamefont {B.}~\bibnamefont {Pannetier}},\ and\ \bibinfo {author} {\bibfnamefont {I.}~\bibnamefont {Protopopov}},\ }\bibfield  {title} {\bibinfo {title} {Measurement of the current-phase relation in josephson junction rhombi chains},\ }\href {https://doi.org/10.1103/PhysRevB.78.104504} {\bibfield  {journal} {\bibinfo  {journal} {Phys. Rev. B}\ }\textbf {\bibinfo {volume} {78}},\ \bibinfo {pages} {104504} (\bibinfo {year} {2008})}\BibitemShut {NoStop}%
\bibitem [{\citenamefont {You}\ \emph {et~al.}(2010)\citenamefont {You}, \citenamefont {Shi}, \citenamefont {Hu},\ and\ \citenamefont {Nori}}]{NoriKitaev2010}%
  \BibitemOpen
  \bibfield  {author} {\bibinfo {author} {\bibfnamefont {J.~Q.}\ \bibnamefont {You}}, \bibinfo {author} {\bibfnamefont {X.-F.}\ \bibnamefont {Shi}}, \bibinfo {author} {\bibfnamefont {X.}~\bibnamefont {Hu}},\ and\ \bibinfo {author} {\bibfnamefont {F.}~\bibnamefont {Nori}},\ }\bibfield  {title} {\bibinfo {title} {Quantum emulation of a spin system with topologically protected ground states using superconducting quantum circuits},\ }\href {https://doi.org/10.1103/PhysRevB.81.014505} {\bibfield  {journal} {\bibinfo  {journal} {Phys. Rev. B}\ }\textbf {\bibinfo {volume} {81}},\ \bibinfo {pages} {014505} (\bibinfo {year} {2010})}\BibitemShut {NoStop}%
\bibitem [{\citenamefont {Johnson}\ \emph {et~al.}(2011)\citenamefont {Johnson}, \citenamefont {Amin}, \citenamefont {Gildert}, \citenamefont {Lanting}, \citenamefont {Hamze}, \citenamefont {Dickson}, \citenamefont {Harris}, \citenamefont {Berkley}, \citenamefont {Johansson}, \citenamefont {Bunyk} \emph {et~al.}}]{johnson2011quantum}%
  \BibitemOpen
  \bibfield  {author} {\bibinfo {author} {\bibfnamefont {M.~W.}\ \bibnamefont {Johnson}}, \bibinfo {author} {\bibfnamefont {M.~H.}\ \bibnamefont {Amin}}, \bibinfo {author} {\bibfnamefont {S.}~\bibnamefont {Gildert}}, \bibinfo {author} {\bibfnamefont {T.}~\bibnamefont {Lanting}}, \bibinfo {author} {\bibfnamefont {F.}~\bibnamefont {Hamze}}, \bibinfo {author} {\bibfnamefont {N.}~\bibnamefont {Dickson}}, \bibinfo {author} {\bibfnamefont {R.}~\bibnamefont {Harris}}, \bibinfo {author} {\bibfnamefont {A.~J.}\ \bibnamefont {Berkley}}, \bibinfo {author} {\bibfnamefont {J.}~\bibnamefont {Johansson}}, \bibinfo {author} {\bibfnamefont {P.}~\bibnamefont {Bunyk}}, \emph {et~al.},\ }\bibfield  {title} {\bibinfo {title} {Quantum annealing with manufactured spins},\ }\href@noop {} {\bibfield  {journal} {\bibinfo  {journal} {Nature}\ }\textbf {\bibinfo {volume} {473}},\ \bibinfo {pages} {194} (\bibinfo {year} {2011})}\BibitemShut {NoStop}%
\bibitem [{\citenamefont {King}\ \emph {et~al.}(2018)\citenamefont {King}, \citenamefont {Carrasquilla}, \citenamefont {Raymond}, \citenamefont {Ozfidan}, \citenamefont {Andriyash}, \citenamefont {Berkley}, \citenamefont {Reis}, \citenamefont {Lanting}, \citenamefont {Harris}, \citenamefont {Altomare} \emph {et~al.}}]{king2018observation}%
  \BibitemOpen
  \bibfield  {author} {\bibinfo {author} {\bibfnamefont {A.~D.}\ \bibnamefont {King}}, \bibinfo {author} {\bibfnamefont {J.}~\bibnamefont {Carrasquilla}}, \bibinfo {author} {\bibfnamefont {J.}~\bibnamefont {Raymond}}, \bibinfo {author} {\bibfnamefont {I.}~\bibnamefont {Ozfidan}}, \bibinfo {author} {\bibfnamefont {E.}~\bibnamefont {Andriyash}}, \bibinfo {author} {\bibfnamefont {A.}~\bibnamefont {Berkley}}, \bibinfo {author} {\bibfnamefont {M.}~\bibnamefont {Reis}}, \bibinfo {author} {\bibfnamefont {T.}~\bibnamefont {Lanting}}, \bibinfo {author} {\bibfnamefont {R.}~\bibnamefont {Harris}}, \bibinfo {author} {\bibfnamefont {F.}~\bibnamefont {Altomare}}, \emph {et~al.},\ }\bibfield  {title} {\bibinfo {title} {Observation of topological phenomena in a programmable lattice of 1,800 qubits},\ }\href@noop {} {\bibfield  {journal} {\bibinfo  {journal} {Nature}\ }\textbf {\bibinfo {volume} {560}},\ \bibinfo {pages} {456} (\bibinfo {year} {2018})}\BibitemShut {NoStop}%
\bibitem [{\citenamefont {King}\ \emph {et~al.}(2021)\citenamefont {King}, \citenamefont {Raymond}, \citenamefont {Lanting}, \citenamefont {Isakov}, \citenamefont {Mohseni}, \citenamefont {Poulin-Lamarre}, \citenamefont {Ejtemaee}, \citenamefont {Bernoudy}, \citenamefont {Ozfidan}, \citenamefont {Smirnov} \emph {et~al.}}]{king2021scaling}%
  \BibitemOpen
  \bibfield  {author} {\bibinfo {author} {\bibfnamefont {A.~D.}\ \bibnamefont {King}}, \bibinfo {author} {\bibfnamefont {J.}~\bibnamefont {Raymond}}, \bibinfo {author} {\bibfnamefont {T.}~\bibnamefont {Lanting}}, \bibinfo {author} {\bibfnamefont {S.~V.}\ \bibnamefont {Isakov}}, \bibinfo {author} {\bibfnamefont {M.}~\bibnamefont {Mohseni}}, \bibinfo {author} {\bibfnamefont {G.}~\bibnamefont {Poulin-Lamarre}}, \bibinfo {author} {\bibfnamefont {S.}~\bibnamefont {Ejtemaee}}, \bibinfo {author} {\bibfnamefont {W.}~\bibnamefont {Bernoudy}}, \bibinfo {author} {\bibfnamefont {I.}~\bibnamefont {Ozfidan}}, \bibinfo {author} {\bibfnamefont {A.~Y.}\ \bibnamefont {Smirnov}}, \emph {et~al.},\ }\bibfield  {title} {\bibinfo {title} {Scaling advantage over path-integral monte carlo in quantum simulation of geometrically frustrated magnets},\ }\href@noop {} {\bibfield  {journal} {\bibinfo  {journal} {Nature communications}\ }\textbf {\bibinfo {volume} {12}},\ \bibinfo {pages} {1113} (\bibinfo {year} {2021})}\BibitemShut
  {NoStop}%
\bibitem [{\citenamefont {Dou{\c{c}}ot}\ and\ \citenamefont {Vidal}(2002)}]{douccot2002pairing}%
  \BibitemOpen
  \bibfield  {author} {\bibinfo {author} {\bibfnamefont {B.}~\bibnamefont {Dou{\c{c}}ot}}\ and\ \bibinfo {author} {\bibfnamefont {J.}~\bibnamefont {Vidal}},\ }\bibfield  {title} {\bibinfo {title} {Pairing of cooper pairs in a fully frustrated josephson-junction chain},\ }\href@noop {} {\bibfield  {journal} {\bibinfo  {journal} {Physical review letters}\ }\textbf {\bibinfo {volume} {88}},\ \bibinfo {pages} {227005} (\bibinfo {year} {2002})}\BibitemShut {NoStop}%
\bibitem [{\citenamefont {Cataudella}\ and\ \citenamefont {Fazio}(2003)}]{cataudella2003glassy}%
  \BibitemOpen
  \bibfield  {author} {\bibinfo {author} {\bibfnamefont {V.}~\bibnamefont {Cataudella}}\ and\ \bibinfo {author} {\bibfnamefont {R.}~\bibnamefont {Fazio}},\ }\bibfield  {title} {\bibinfo {title} {Glassy dynamics of josephson arrays on a dice lattice},\ }\href {http://stacks.iop.org/0295-5075/61/i=3/a=341} {\bibfield  {journal} {\bibinfo  {journal} {EPL (Europhysics Letters)}\ }\textbf {\bibinfo {volume} {61}},\ \bibinfo {pages} {341} (\bibinfo {year} {2003})}\BibitemShut {NoStop}%
\bibitem [{\citenamefont {Daley}\ \emph {et~al.}(2022)\citenamefont {Daley}, \citenamefont {Bloch}, \citenamefont {Kokail}, \citenamefont {Flannigan}, \citenamefont {Pearson}, \citenamefont {Troyer},\ and\ \citenamefont {Zoller}}]{daley2022practical}%
  \BibitemOpen
  \bibfield  {author} {\bibinfo {author} {\bibfnamefont {A.~J.}\ \bibnamefont {Daley}}, \bibinfo {author} {\bibfnamefont {I.}~\bibnamefont {Bloch}}, \bibinfo {author} {\bibfnamefont {C.}~\bibnamefont {Kokail}}, \bibinfo {author} {\bibfnamefont {S.}~\bibnamefont {Flannigan}}, \bibinfo {author} {\bibfnamefont {N.}~\bibnamefont {Pearson}}, \bibinfo {author} {\bibfnamefont {M.}~\bibnamefont {Troyer}},\ and\ \bibinfo {author} {\bibfnamefont {P.}~\bibnamefont {Zoller}},\ }\bibfield  {title} {\bibinfo {title} {Practical quantum advantage in quantum simulation},\ }\href@noop {} {\bibfield  {journal} {\bibinfo  {journal} {Nature}\ }\textbf {\bibinfo {volume} {607}},\ \bibinfo {pages} {667} (\bibinfo {year} {2022})}\BibitemShut {NoStop}%
\bibitem [{\citenamefont {Buluta}\ and\ \citenamefont {Nori}(2009)}]{buluta2009quantum}%
  \BibitemOpen
  \bibfield  {author} {\bibinfo {author} {\bibfnamefont {I.}~\bibnamefont {Buluta}}\ and\ \bibinfo {author} {\bibfnamefont {F.}~\bibnamefont {Nori}},\ }\bibfield  {title} {\bibinfo {title} {Quantum simulators},\ }\href@noop {} {\bibfield  {journal} {\bibinfo  {journal} {Science}\ }\textbf {\bibinfo {volume} {326}},\ \bibinfo {pages} {108} (\bibinfo {year} {2009})}\BibitemShut {NoStop}%
\bibitem [{\citenamefont {Caputo}\ \emph {et~al.}(2001)\citenamefont {Caputo}, \citenamefont {Fistul},\ and\ \citenamefont {Ustinov}}]{caputo2001resonances}%
  \BibitemOpen
  \bibfield  {author} {\bibinfo {author} {\bibfnamefont {P.}~\bibnamefont {Caputo}}, \bibinfo {author} {\bibfnamefont {M.}~\bibnamefont {Fistul}},\ and\ \bibinfo {author} {\bibfnamefont {A.}~\bibnamefont {Ustinov}},\ }\bibfield  {title} {\bibinfo {title} {Resonances in one and two rows of triangular josephson junction cells},\ }\href@noop {} {\bibfield  {journal} {\bibinfo  {journal} {Physical Review B}\ }\textbf {\bibinfo {volume} {63}},\ \bibinfo {pages} {214510} (\bibinfo {year} {2001})}\BibitemShut {NoStop}%
\bibitem [{\citenamefont {Valdez-Balderas}\ and\ \citenamefont {Stroud}(2005)}]{valdez2005superconductivity}%
  \BibitemOpen
  \bibfield  {author} {\bibinfo {author} {\bibfnamefont {D.}~\bibnamefont {Valdez-Balderas}}\ and\ \bibinfo {author} {\bibfnamefont {D.}~\bibnamefont {Stroud}},\ }\bibfield  {title} {\bibinfo {title} {Superconductivity versus phase separation, stripes, and checkerboard ordering: A two-dimensional monte carlo study},\ }\href@noop {} {\bibfield  {journal} {\bibinfo  {journal} {Physical Review B}\ }\textbf {\bibinfo {volume} {72}},\ \bibinfo {pages} {214501} (\bibinfo {year} {2005})}\BibitemShut {NoStop}%
\bibitem [{\citenamefont {Rizzi}\ \emph {et~al.}(2006)\citenamefont {Rizzi}, \citenamefont {Cataudella},\ and\ \citenamefont {Fazio}}]{rizzi20064}%
  \BibitemOpen
  \bibfield  {author} {\bibinfo {author} {\bibfnamefont {M.}~\bibnamefont {Rizzi}}, \bibinfo {author} {\bibfnamefont {V.}~\bibnamefont {Cataudella}},\ and\ \bibinfo {author} {\bibfnamefont {R.}~\bibnamefont {Fazio}},\ }\bibfield  {title} {\bibinfo {title} {4 e-condensation in a fully frustrated josephson junction diamond chain},\ }\href@noop {} {\bibfield  {journal} {\bibinfo  {journal} {Physical Review B}\ }\textbf {\bibinfo {volume} {73}},\ \bibinfo {pages} {100502} (\bibinfo {year} {2006})}\BibitemShut {NoStop}%
\bibitem [{\citenamefont {Andreanov}\ and\ \citenamefont {Fistul}(2019)}]{andreanov2019resonant}%
  \BibitemOpen
  \bibfield  {author} {\bibinfo {author} {\bibfnamefont {A.}~\bibnamefont {Andreanov}}\ and\ \bibinfo {author} {\bibfnamefont {M.}~\bibnamefont {Fistul}},\ }\bibfield  {title} {\bibinfo {title} {Resonant frequencies and spatial correlations in frustrated arrays of josephson type nonlinear oscillators},\ }\href@noop {} {\bibfield  {journal} {\bibinfo  {journal} {Journal of Physics A: Mathematical and Theoretical}\ }\textbf {\bibinfo {volume} {52}},\ \bibinfo {pages} {105101} (\bibinfo {year} {2019})}\BibitemShut {NoStop}%
\bibitem [{\citenamefont {Andreanov}\ and\ \citenamefont {Fistul}(2020)}]{andreanov2020frustration}%
  \BibitemOpen
  \bibfield  {author} {\bibinfo {author} {\bibfnamefont {A.}~\bibnamefont {Andreanov}}\ and\ \bibinfo {author} {\bibfnamefont {M.}~\bibnamefont {Fistul}},\ }\bibfield  {title} {\bibinfo {title} {Frustration-induced highly anisotropic magnetic patterns in the classical xy model on the kagome lattice},\ }\href@noop {} {\bibfield  {journal} {\bibinfo  {journal} {Physical Review B}\ }\textbf {\bibinfo {volume} {102}},\ \bibinfo {pages} {140405} (\bibinfo {year} {2020})}\BibitemShut {NoStop}%
\bibitem [{\citenamefont {Orlando}\ \emph {et~al.}(1999)\citenamefont {Orlando}, \citenamefont {Mooij}, \citenamefont {Tian}, \citenamefont {Van Der~Wal}, \citenamefont {Levitov}, \citenamefont {Lloyd},\ and\ \citenamefont {Mazo}}]{orlando1999superconducting}%
  \BibitemOpen
  \bibfield  {author} {\bibinfo {author} {\bibfnamefont {T.}~\bibnamefont {Orlando}}, \bibinfo {author} {\bibfnamefont {J.}~\bibnamefont {Mooij}}, \bibinfo {author} {\bibfnamefont {L.}~\bibnamefont {Tian}}, \bibinfo {author} {\bibfnamefont {C.~H.}\ \bibnamefont {Van Der~Wal}}, \bibinfo {author} {\bibfnamefont {L.}~\bibnamefont {Levitov}}, \bibinfo {author} {\bibfnamefont {S.}~\bibnamefont {Lloyd}},\ and\ \bibinfo {author} {\bibfnamefont {J.}~\bibnamefont {Mazo}},\ }\bibfield  {title} {\bibinfo {title} {Superconducting persistent-current qubit},\ }\href@noop {} {\bibfield  {journal} {\bibinfo  {journal} {Physical Review B}\ }\textbf {\bibinfo {volume} {60}},\ \bibinfo {pages} {15398} (\bibinfo {year} {1999})}\BibitemShut {NoStop}%
\bibitem [{\citenamefont {Jung}\ \emph {et~al.}(2014)\citenamefont {Jung}, \citenamefont {Butz}, \citenamefont {Marthaler}, \citenamefont {Fistul}, \citenamefont {Lepp{\"a}kangas}, \citenamefont {Koshelets},\ and\ \citenamefont {Ustinov}}]{jung2014multistability}%
  \BibitemOpen
  \bibfield  {author} {\bibinfo {author} {\bibfnamefont {P.}~\bibnamefont {Jung}}, \bibinfo {author} {\bibfnamefont {S.}~\bibnamefont {Butz}}, \bibinfo {author} {\bibfnamefont {M.}~\bibnamefont {Marthaler}}, \bibinfo {author} {\bibfnamefont {M.~V.}\ \bibnamefont {Fistul}}, \bibinfo {author} {\bibfnamefont {J.}~\bibnamefont {Lepp{\"a}kangas}}, \bibinfo {author} {\bibfnamefont {V.~P.}\ \bibnamefont {Koshelets}},\ and\ \bibinfo {author} {\bibfnamefont {A.~V.}\ \bibnamefont {Ustinov}},\ }\bibfield  {title} {\bibinfo {title} {Multistability and switching in a superconducting metamaterial},\ }\href {http://dx.doi.org/10.1038/ncomms4730} {\bibfield  {journal} {\bibinfo  {journal} {Nat. Comm.}\ }\textbf {\bibinfo {volume} {5}},\ \bibinfo {pages} {3730 EP } (\bibinfo {year} {2014})}\BibitemShut {NoStop}%
\bibitem [{\citenamefont {Shulga}\ \emph {et~al.}(2018)\citenamefont {Shulga}, \citenamefont {Il’ichev}, \citenamefont {Fistul}, \citenamefont {Besedin}, \citenamefont {Butz}, \citenamefont {Astafiev}, \citenamefont {H{\"u}bner},\ and\ \citenamefont {Ustinov}}]{shulga2018magnetically}%
  \BibitemOpen
  \bibfield  {author} {\bibinfo {author} {\bibfnamefont {K.}~\bibnamefont {Shulga}}, \bibinfo {author} {\bibfnamefont {E.}~\bibnamefont {Il’ichev}}, \bibinfo {author} {\bibfnamefont {M.~V.}\ \bibnamefont {Fistul}}, \bibinfo {author} {\bibfnamefont {I.}~\bibnamefont {Besedin}}, \bibinfo {author} {\bibfnamefont {S.}~\bibnamefont {Butz}}, \bibinfo {author} {\bibfnamefont {O.}~\bibnamefont {Astafiev}}, \bibinfo {author} {\bibfnamefont {U.}~\bibnamefont {H{\"u}bner}},\ and\ \bibinfo {author} {\bibfnamefont {A.~V.}\ \bibnamefont {Ustinov}},\ }\bibfield  {title} {\bibinfo {title} {Magnetically induced transparency of a quantum metamaterial composed of twin flux qubits},\ }\href@noop {} {\bibfield  {journal} {\bibinfo  {journal} {Nature communications}\ }\textbf {\bibinfo {volume} {9}},\ \bibinfo {pages} {150} (\bibinfo {year} {2018})}\BibitemShut {NoStop}%
\bibitem [{\citenamefont {Feofanov}\ \emph {et~al.}(2010)\citenamefont {Feofanov}, \citenamefont {Oboznov}, \citenamefont {Bol'ginov}, \citenamefont {Lisenfeld}, \citenamefont {Poletto}, \citenamefont {Ryazanov}, \citenamefont {Rossolenko}, \citenamefont {Khabipov}, \citenamefont {Balashov}, \citenamefont {Zorin}, \citenamefont {Dmitriev}, \citenamefont {Koshelets},\ and\ \citenamefont {Ustinov}}]{feofanov2010implementation}%
  \BibitemOpen
  \bibfield  {author} {\bibinfo {author} {\bibfnamefont {A.~K.}\ \bibnamefont {Feofanov}}, \bibinfo {author} {\bibfnamefont {V.~A.}\ \bibnamefont {Oboznov}}, \bibinfo {author} {\bibfnamefont {V.~V.}\ \bibnamefont {Bol'ginov}}, \bibinfo {author} {\bibfnamefont {J.}~\bibnamefont {Lisenfeld}}, \bibinfo {author} {\bibfnamefont {S.}~\bibnamefont {Poletto}}, \bibinfo {author} {\bibfnamefont {V.~V.}\ \bibnamefont {Ryazanov}}, \bibinfo {author} {\bibfnamefont {A.~N.}\ \bibnamefont {Rossolenko}}, \bibinfo {author} {\bibfnamefont {M.}~\bibnamefont {Khabipov}}, \bibinfo {author} {\bibfnamefont {D.}~\bibnamefont {Balashov}}, \bibinfo {author} {\bibfnamefont {A.~B.}\ \bibnamefont {Zorin}}, \bibinfo {author} {\bibfnamefont {P.~N.}\ \bibnamefont {Dmitriev}}, \bibinfo {author} {\bibfnamefont {V.~P.}\ \bibnamefont {Koshelets}},\ and\ \bibinfo {author} {\bibfnamefont {A.~V.}\ \bibnamefont {Ustinov}},\ }\bibfield  {title} {\bibinfo {title} {Implementation of superconductor/ferromagnet/ superconductor $\pi$-shifters in
  superconducting digital and quantum circuits},\ }\href {http://dx.doi.org/10.1038/nphys1700} {\bibfield  {journal} {\bibinfo  {journal} {Nat. Phys.}\ }\textbf {\bibinfo {volume} {6}},\ \bibinfo {pages} {593 EP } (\bibinfo {year} {2010})}\BibitemShut {NoStop}%
\bibitem [{\citenamefont {Hilgenkamp}(2008)}]{hilgenkamp2008pi}%
  \BibitemOpen
  \bibfield  {author} {\bibinfo {author} {\bibfnamefont {H.}~\bibnamefont {Hilgenkamp}},\ }\bibfield  {title} {\bibinfo {title} {Pi-phase shift josephson structures},\ }\href {http://stacks.iop.org/0953-2048/21/i=3/a=034011} {\bibfield  {journal} {\bibinfo  {journal} {Supercond. Sci. and Tech.}\ }\textbf {\bibinfo {volume} {21}},\ \bibinfo {pages} {034011} (\bibinfo {year} {2008})}\BibitemShut {NoStop}%
\bibitem [{\citenamefont {Dias}\ and\ \citenamefont {Marques}(2011)}]{dias2011frustrated}%
  \BibitemOpen
  \bibfield  {author} {\bibinfo {author} {\bibfnamefont {R.}~\bibnamefont {Dias}}\ and\ \bibinfo {author} {\bibfnamefont {A.}~\bibnamefont {Marques}},\ }\bibfield  {title} {\bibinfo {title} {Frustrated multiband superconductivity},\ }\href@noop {} {\bibfield  {journal} {\bibinfo  {journal} {Superconductor Science and Technology}\ }\textbf {\bibinfo {volume} {24}},\ \bibinfo {pages} {085009} (\bibinfo {year} {2011})}\BibitemShut {NoStop}%
\bibitem [{\citenamefont {Lucas}(2014)}]{lucas2014ising}%
  \BibitemOpen
  \bibfield  {author} {\bibinfo {author} {\bibfnamefont {A.}~\bibnamefont {Lucas}},\ }\bibfield  {title} {\bibinfo {title} {Ising formulations of many np problems},\ }\href@noop {} {\bibfield  {journal} {\bibinfo  {journal} {Frontiers in physics}\ }\textbf {\bibinfo {volume} {2}},\ \bibinfo {pages} {5} (\bibinfo {year} {2014})}\BibitemShut {NoStop}%
\end{thebibliography}%

\end{document}